\documentclass[usenatbib,onecolumn]{aastex62}
\usepackage{graphicx}
\usepackage{amsmath,amsfonts,amssymb}

\usepackage{subfigure}

\usepackage{pstricks} 

\submitjournal{ApJ}

\shorttitle{dynamical effects in period derivatives}
\shortauthors{Pathak and Bagchi}

\begin{document}

\title[dynamical effects in period derivatives]{Dynamical effects in the observed rate of change of the orbital and the spin periods of radio pulsars: Improvement in the method of estimation and its implications}

\correspondingauthor{Dhruv Pathak}
\email{dhruvpathak@imsc.res.in, pathakdhruv9786@gmail.com}

\author[0000-0001-8129-0473]{Dhruv Pathak}
\affil{The Institute of Mathematical Sciences\\ 
C. I. T. campus, Taramani, Chennai, 600113, India}
\affiliation{Homi Bhabha National Institute\\
Training School Complex, Anushakti Nagar, Mumbai 400094, India}

\author[0000-0001-8640-8186]{Manjari Bagchi}
\affil{The Institute of Mathematical Sciences\\ 
C. I. T. campus, Taramani, Chennai, 600113, India}
\affiliation{Homi Bhabha National Institute\\
Training School Complex, Anushakti Nagar, Mumbai 400094, India}

\begin{abstract}
The observed values of the rate of change of the orbital and the spin periods of pulsars are affected by different dynamical effects, for example, the line-of-sight acceleration and the proper motion of the pulsar relative to the sun. We explore these dynamical effects thoroughly and point out the drawbacks of popular methods. We introduce a package, `GalDynPsr', that evaluates different dynamical effects following traditional as well as improved methods based on the model of the Galactic potential provided in a publicly available package called `galpy'. We argue that the improved methods introduced in this paper should be used for pulsars located 1 kpc or farther away from the solar system, especially when precise values of the rate of change of the periods are required, e.g., while placing limits on alternative theories of gravity, calculating the spin-down limit of the continuous gravitational waves emitted from a rotationally deformed neutron star, understanding pulsar `death-line', etc. GalDynPsr is available online and open for contributions.
\end{abstract}


\section{Introduction} \label{sec:intro}

Precision timing analysis produces many interesting results, examples include the discovery of a millisecond pulsar in a stellar triple system \citep{rsa14} and the test of the universality of free fall with that pulsar \citep{agh18}, the discovery of a millisecond pulsar in an extremely wide binary \citep{bjs16, kp16}, the evidence of an intermediate-mass black hole at the center of the globular cluster NGC 6624 \citep{bsl17}, etc. We might expect many such discoveries in the future as a good number of stable millisecond pulsars are being timed regularly for long time-baselines, and more pulsars are expected to be added in this list when the Square Kilometer Array (SKA) will become operative. Such precision timing experiments are expected to contribute in many areas of fundamental physics, including the detection of the low-frequency gravitational waves, improvements in the tests of gravity theory, stricter constraints on the dense matter equation of state, better understanding of the spin-down and the spin-up related phenomena, and so on \citep{kra16}.

In this connection, it is important to understand external factors affecting the values of the pulsar and the orbital parameters estimated in the timing analysis, especially when one aims to achieve goals like testing theories of gravity or precise measurements of spin-down related phenomena.

Among various possible external factors, we presently aim to quantify the dynamical effects as precisely as possible. We discuss how these dynamical effects can manifest in the values of the rate of change of the orbital and the spin periods and how these effects are accounted for. We point out the limitations of existing methods and argue the need of improved methods to extract these effects. We have created a package  `GalDynPsr' for this purpose. In the present paper, we discuss salient features of GalDynPsr with some examples and the potential roles it can play while interpreting results of precise timing experiments. In fact, GalDynPsr has been already used by \citet{agh18} to place the best ever limit of the non-violation of the universality of free fall. GalDynPsr can be used in many other similarly important studies, some of which we have discussed in this paper.

\section{Basics of dynamical effects}
\label{sec:dyneffects}

Pulsar timing enables us to measure various parameters of the pulsar like its spin period, the rate of change of the spin period, its position, proper motion, etc., as well as various Keplerian and post-Keplerian orbital parameters if the pulsar is in a binary. The orbital period is one of the Keplerian parameters while the rate of change of the orbital period is one of the post-Keplerian parameters. Along with other parameters, periods and period derivatives reveal properties of pulsars, their evolutionary histories, and assist to test various postulates of fundamental physics. However, to achive these goals, it is essential to estimate the `true' or `intrinsic' values of  these parameters precisely by understanding and eliminating external factors as accurately as possible. For a clean binary system, i.e., where gravity is the only interaction between the pulsar and its companion, it is believed that the orbital period decreases due to the radiation of the gravitational waves\footnote{Otherwise, various other phenomena might cause the orbital period to change, e.g., the mass loss from the binary, tidal interactions between the members of the binary, etc. If exist, these phenomena would keep their imprints on the evolution and properties of binaries. So, it would be worthwhile to quantify these effects. Elimination of external effects as discussed in this paper is essential to achieve this goal.}, and hence the rate of change of the orbital period must be negative. On the other hand, the existence of various forms of energy losses from pulsars, e.g., the emission of electromagnetic waves, the emission of winds of charged particles, the expected emission of continuous gravitational waves due to the rotational deformation of the pulsar, etc. lead to the expectation of the rate of change of the spin-periods of non-accreting rotation powered radio pulsars to be positive. The signs of the measured rate of change of the orbital and the spin periods usually comply with these expectations although there are a few of exceptions for both of the parameters. These exceptions are explained as results of external effects, some of which are subject of discussions of the present paper. Although the values of the rate of change of the spin and the orbital periods are very small, these can be measured accurately if a timing campaign is pursued for long enough.  As an example, 15-years of data of PSR J0437$-$4715 by the Parkes Pulsar Timing Array enabled the measurements of both of the period derivatives very accurately, e.g., the rate of change of the spin frequency\footnote{It is conventional to fit for spin frequency and its derivative instead of spin period and its derivative.} is $-1.728361(5) \times 10^{-15}$ and the rate of change of the orbital period as $3.728(6) \times 10^{-12}$ \citep{rhc16}. Such accurate measurements would be useless if the knowledge of the external factors are not accurate enough. The Doppler shift is one of these external factors. Like any other periodic functions, both the spin period and the orbital period experience the Doppler shift due to the relative motion between the pulsar (the source) and the observer. Here, we discuss this effect and methods to eliminate it.

During the procedure of pulsar timing analysis, the site arrival times are translated to the barycentric arrival times. As the barycenter of the solar system lies very close to the surface of the sun (within 0.008 AU from the center of the sun \citep{ps11}), in the present paper we use the word `sun' in the place of `the barycenter'. So, the motion between the sun and the pulsar comes into play in the expression of the Doppler shift, and the observed value of the orbital period (${P}_{\rm b, obs}$) of the pulsar is related to its intrinsic value (${P}_{\rm b, int}$) as:

\begin{equation}
P_{\rm b, int} = (c + \vec{V}_{\rm s} . \widehat{n}_{\rm s p}) (c + \vec{V}_{\rm p} . \widehat{n}_{\rm s p})^{-1} P_{\rm b, obs} ~,
\label{eq:doppler1}
\end{equation} where $\vec{V}_{\rm s}$ is the velocity of the sun, $ \vec{V}_{\rm p} $ is the velocity of the pulsar, $ \widehat{n}_{\rm s p}$ is a unit vector from the sun to the pulsar, and c is the speed of light. As $1 + \frac{\vec{V}_s . \widehat{n}_{\rm s p}}{c} \approx 1$ and $1 + \frac{\vec{V}_p . \widehat{n}_{\rm s p}}{c} \approx 1$, $P_{\rm b, int} \simeq P_{\rm b, obs}$ and can be written as $P_{\rm b}$ for the sake of simplicity. However, Eqn. (\ref{eq:doppler1}) has an interesting implication on the observed rate of change of the orbital period. Differentiating Eqn. (\ref{eq:doppler1}) with respect to time and dividing by Eqn. (\ref{eq:doppler1}), we get:

\begin{equation}
\frac{\dot{P}_{\rm b, int} }{ {P}_{\rm b}} = \frac{ (\vec{a}_s - \vec{a}_p) \cdot \widehat{n}_{\rm s p}}{c} + \frac{1}{c} (\vec{V}_s - \vec{V}_p) \cdot \frac{d}{dt} (\widehat{n}_{\rm s p})+ \frac{\dot{P}_{\rm b, obs} }{ {P}_{\rm b}}  ~,
\label{eq:doppler2}
\end{equation} where $\vec{a}_s = d \vec{V}_s / dt$ is the acceleration of the sun and $\vec{a}_p =d \vec{V}_p / dt$ is the acceleration of the pulsar. The gradient of the gravitational potential of the Galaxy is the main source of accelerations of the objects in it. There can be additional sources, like local potentials, orbital motions of the pulsars, etc. Note that, the first and the second terms of the above equation are small, but usually within a few orders of magnitude of $\dot{P}_{\rm b, obs} / {P}_{\rm b}$, and hence cannot be neglected. Eqn. \ref{eq:doppler2} can be rewritten as:

\begin{equation}
\left( \frac{\dot{P}_{\rm b} }{ {P}_{\rm b}} \right)_{\rm excess} =  \frac{\dot{P}_{\rm b, obs} }{ {P}_{\rm b}}   -  \frac{\dot{P}_{\rm b, int} }{ {P}_{\rm b}}   = - \left[ \frac{ (\vec{a}_s - \vec{a}_p) \cdot \widehat{n}_{\rm s p}}{c} + \frac{1}{c} (\vec{V}_s - \vec{V}_p) \cdot \frac{d}{dt} (\widehat{n}_{\rm s p}) \right] ~.
\label{eq:doppler2b}
\end{equation}

The first term in the right hand side of Eqn. (\ref{eq:doppler2b}) depends on the relative acceleration that can be decomposed into two components, one is parallel to the Galactic plane and the other is perpendicular to the Galactic plane. These two components are denoted by $\left( \frac{\dot{P}_{\rm b} }{ {P}_{\rm b}} \right)_{\rm excess, Galpl}$ and $ \left( \frac{\dot{P}_{\rm b} }{ {P}_{\rm b}} \right)_{\rm excess, Galz}$, respectively, assuming that those are caused by the gravitational potential of the Galaxy. The second term in the right hand side of Eqn. (\ref{eq:doppler2b}), which involves the relative velocity and the change of the location of the pulsar, is the well-known `Shklovskii term' \citep{shk70} and is denoted by $\left( \frac{\dot{P}_{\rm b} }{ {P}_{\rm b}} \right)_{\rm excess, Shk}$. So, we can write

\begin{equation}
\left( \frac{\dot{P}_{\rm b} }{ {P}_{\rm b}} \right)_{\rm excess} = \left( \frac{\dot{P}_{\rm b} }{ {P}_{\rm b}} \right)_{\rm excess, Galpl}  + \left( \frac{\dot{P}_{\rm b} }{ {P}_{\rm b}} \right)_{\rm excess, Galz}  + \left( \frac{\dot{P}_{\rm b} }{ {P}_{\rm b}} \right)_{\rm excess, Shk} ~.
\label{eq:doppler3}
\end{equation}

The terms in the right hand side of Eqn. (\ref{eq:doppler3}) can be calculated using a model of the gravitational potential of the Galaxy and the measured values of the locations and the motions of the pulsars. 

The fractional dynamical terms shown in Eqn. (\ref{eq:doppler3}) can be converted to the absolute dynamical terms as: $\dot{P}_{\rm b, Galpl} = {P}_{\rm b} \, \left( \frac{\dot{P}_{\rm b} }{ {P}_{\rm b}} \right)_{\rm excess, Galpl}$, $\dot{P}_{\rm b, Galz} = {P}_{\rm b} \, \left( \frac{\dot{P}_{\rm b} }{ {P}_{\rm b}} \right)_{\rm excess, Galz}$, $\dot{P}_{\rm b, Gal} = \dot{P}_{\rm b, Galpl} + \dot{P}_{\rm b, Galz}$, and $\dot{P}_{\rm b, Shk} = {P}_{\rm b} \, \left( \frac{\dot{P}_{\rm b} }{ {P}_{\rm b}} \right)_{\rm excess, Shk} $. So, $\dot{P}_{\rm b, int} = \dot{P}_{\rm b, obs} - \dot{P}_{\rm b, Gal} - \dot{P}_{\rm b, Shk}$.

Similar expressions can be written involving the spin period ($P_{\rm s}$) and the observed and the intrinsic values of the rate of change of the spin period ($\dot{P}_{\rm s, obs}$ and $\dot{P}_{\rm s, int}$), leading to $\left( \frac{\dot{P}_{\rm b} }{ {P}_{\rm b}} \right)_{\rm excess, Galpl} = \left( \frac{\dot{P}_{\rm s} }{ {P}_{\rm s}} \right)_{\rm excess, Galpl} $, $\left( \frac{\dot{P}_{\rm b} }{ {P}_{\rm b}} \right)_{\rm excess, Galz} = \left( \frac{\dot{P}_{\rm s} }{ {P}_{\rm s}} \right)_{\rm excess, Galz} $, $\left( \frac{\dot{P}_{\rm b} }{ {P}_{\rm b}} \right)_{\rm excess, Shk} = \left( \frac{\dot{P}_{\rm s} }{ {P}_{\rm s}} \right)_{\rm excess, Shk} $, and $\left( \frac{\dot{P}_{\rm b} }{ {P}_{\rm b}} \right)_{\rm excess} = \left( \frac{\dot{P}_{\rm s} }{ {P}_{\rm s}} \right)_{\rm excess} $. Absolute dynamical terms in the rate of change of the spin period can be calculated by multiplying these fractional dynamical terms by ${P}_{\rm s}$. However, if the pulsar is in an unmodeled binary, the expression for $\left( \frac{\dot{P}_{\rm s} }{ {P}_{\rm s}} \right)_{\rm excess}$ will have an extra term  caused by the line-of-sight component of the orbital acceleration. There will be additional terms in both $\left( \frac{\dot{P}_{\rm b} }{ {P}_{\rm b}} \right)_{\rm excess}$ and $\left( \frac{\dot{P}_{\rm s} }{ {P}_{\rm s}} \right)_{\rm excess}$ if the pulsar is in a globular cluster. Although, these two additional effects are discussed in the present paper, one needs to calculate those on a case-by-case basis and it is not possible to create a generalized code. There might be other dynamical effects arising due to the existence of nearby massive stars, giant molecular clouds, local stellar inhomogeneity, etc \citep[see section III]{dt91}. We do not discuss these effects further.

Successive differentiation of Eqn. (\ref{eq:doppler1}) will lead to expressions of $\left( \frac{\ddot{P}_{\rm b} }{ {P}_{\rm b}} \right)_{\rm excess} = \left( \frac{\ddot{P}_{\rm s} }{ {P}_{\rm s}} \right)_{\rm excess}$, $\left( \frac{\dddot{P}_{\rm b} }{ {P}_{\rm b}} \right)_{\rm excess} = \left( \frac{\dddot{P}_{\rm s} }{ {P}_{\rm s}} \right)_{\rm excess}$, and so on \citep[and references therein]{lbs18}. These expressions will involve derivatives of relative accelerations like $\frac{1}{c} \frac{d}{dt} (\vec{a}_s - \vec{a}_p) \cdot \widehat{n}_{\rm s p}$, $\frac{1}{c}  \frac{d^2}{dt^2} (\vec{a}_s - \vec{a}_p) \cdot \widehat{n}_{\rm s p}$, etc. plus additional terms that are usually infinitesimal. Although generally small, $\frac{1}{c} \frac{d}{dt} (\vec{a}_s - \vec{a}_p) \cdot \widehat{n}_{\rm s p}$, $\frac{1}{c}  \frac{d^2}{dt^2} (\vec{a}_s - \vec{a}_p) \cdot \widehat{n}_{\rm s p}$, etc. might be non-negligible if they are due to the orbital motion of an unmodeled binary and can eventually help constrain the parameters of the binary \citep{jr97, bjs16, bsl17}. These terms will also be significant in a region where the gradient of the gravitational potential changes rapidly with the spatial coordinates, i.e., inside globular clusters or close to the Galactic center. In those regions, a slight movement of the pulsar from one position to another would yield a change in its acceleration. Moreover, in such dense stellar environments, the probability of close fly-bys causing an impulsive change in the acceleration of the pulsar, is also high. However, higher derivatives of the spin period in the timing solution might arise due to the intrinsic `timing noise' too. Careful analysis and the use of good noise models \citep{sr10, chc11} can help us overcome such ambiguities.

\subsection{Dynamical effects due to the acceleration of the pulsar relative to the sun: A review of traditional methods}
\label{sec:dyneffectstraditional}

We have already mentioned that the first term in the right hand side of Eqn. (\ref{eq:doppler2b}) depends on the relative acceleration that can be decomposed into two components, one is parallel to the Galactic plane and the other is perpendicular to the Galactic plane. Conventionally, instead of a model of the gravitational potential of the Galaxy, people adopt different approximations to evaluate these terms. As these approximations are still in practice, we first discuss these and point out their limitations and mention how we overcome these limitations using a model of the gravitational potential of the Galaxy. The use of this model potential will be elaborated later. We also explain the method to evaluate the second term in the right hand side of Eqn. (\ref{eq:doppler2b}) and discuss extra contributions in the first term in the right hand side of Eqn. (\ref{eq:doppler2b}) that arise in special situations.

\subsubsection{Contribution from the relative acceleration parallel to the Galactic plane}
\label{subsec:galpl}

\begin{figure*}
 \begin{center}
\hskip -1.0cm \subfigure[]{\label{subfig:schematic1}\includegraphics[width=0.5\textwidth,angle=0]{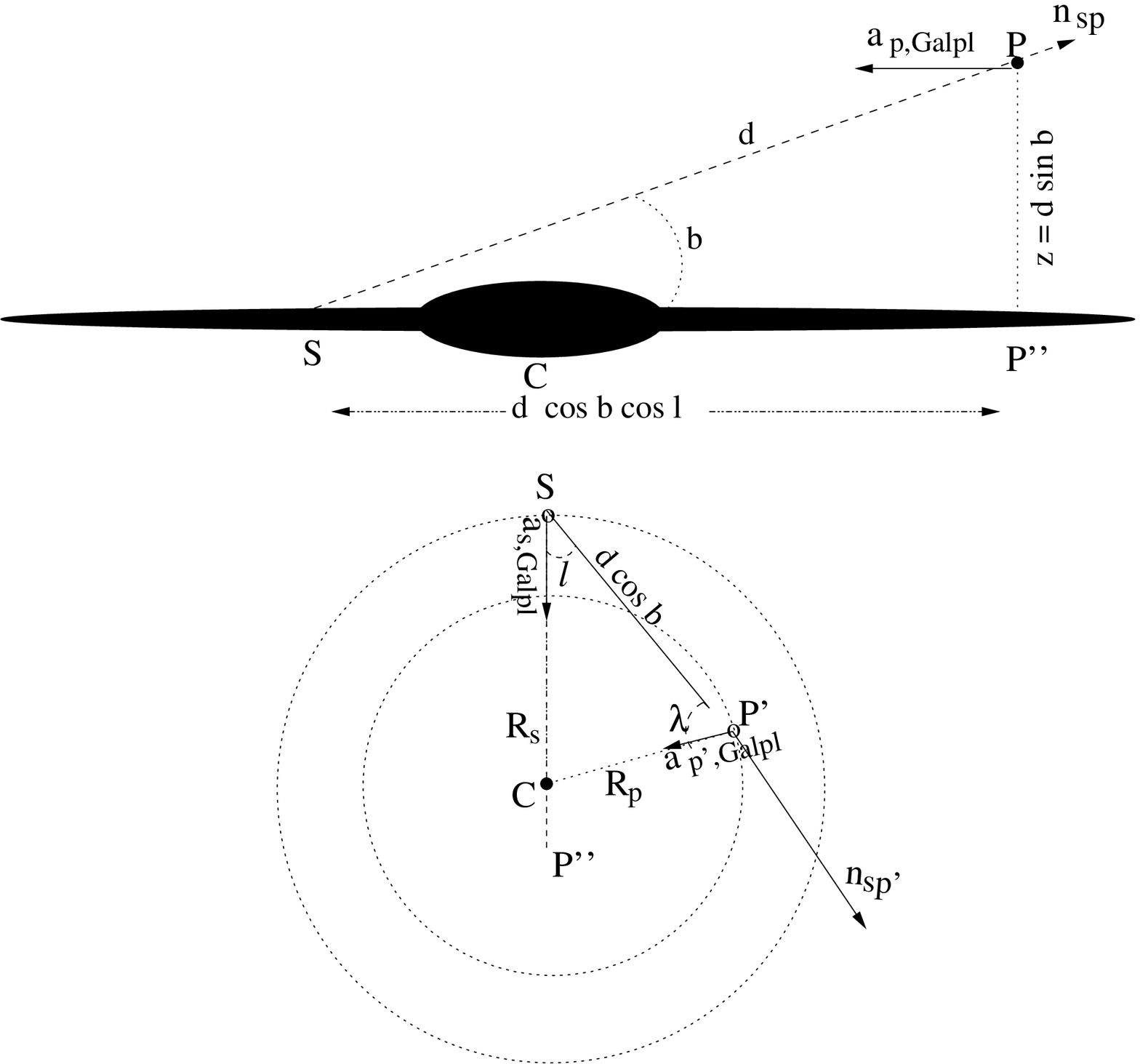}}
 \subfigure[]{\label{subfig:schematic2}\includegraphics[width=0.5\textwidth,angle=0]{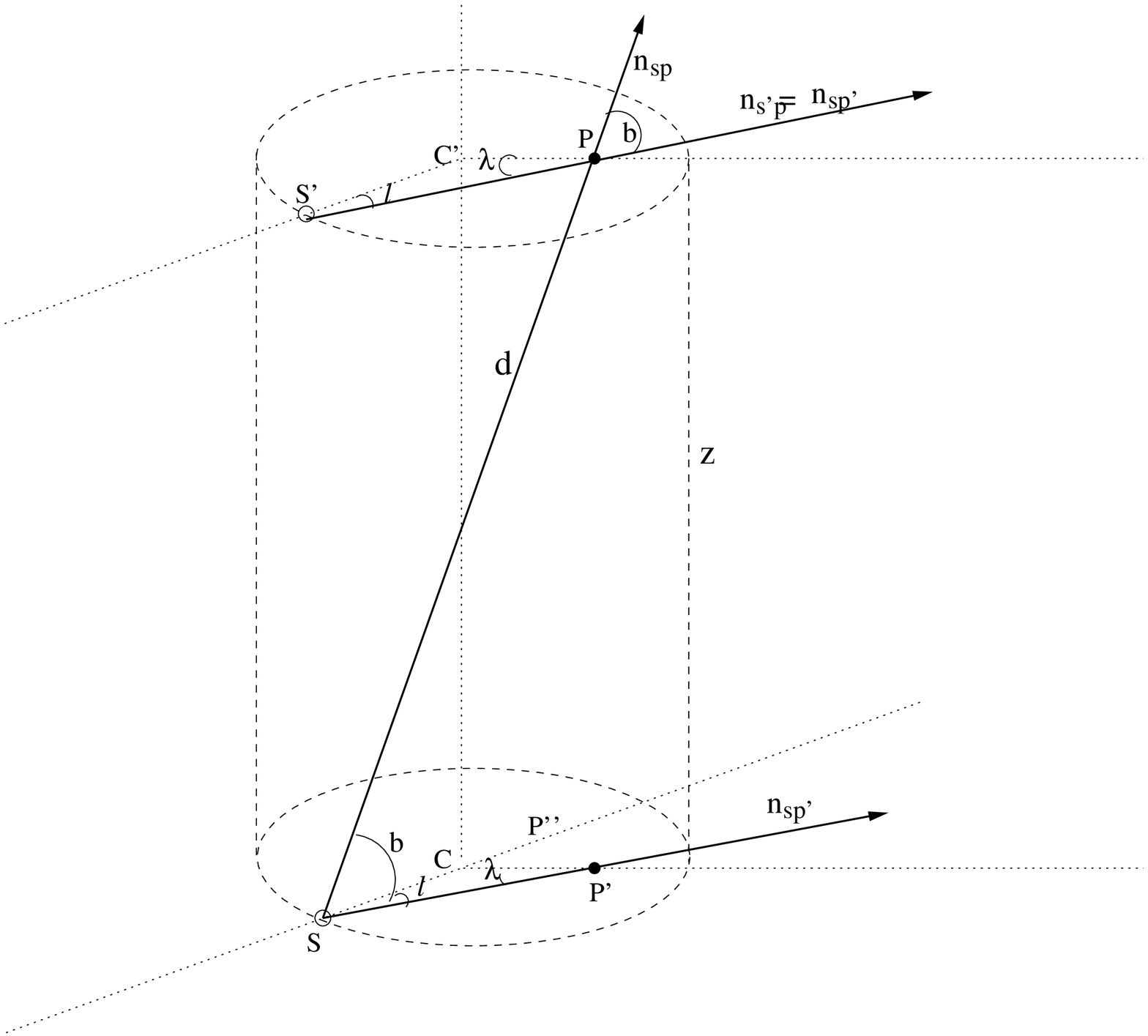}}
 \end{center}
\caption{The schematic diagram to understand accelerations parallel to the Galactic plane. At the left (panel-a), the edge-on view is displayed at the top and the face-on view at the bottom. The right panel (panel-b) shows the three-dimensional geometry. In both of the panels, S is the location of the sun, P is the location of the pulsar, ${\rm P^{\prime}}$ is the projection of P on the Galactic plane, C is the Galactic center, and ${\rm P^{\prime \prime}}$ is the projection of ${\rm P^{\prime}}$ on SC. The longitude, the latitude, and the distance of the pulsar are $l$, $b$, and $d$ respectively. In panel-b, ${\rm S}^{\prime}$ is the projection of S on a plane parallel to the Galactic plane at a height $z$ of the pulsar and ${\rm C}^{\prime}$ is the projection of C on the same plane. This panel also shows three unit vectors, $\widehat{n}_{\rm s p}$ from S to P, $\widehat{n}_{\rm s p^{\prime}}$ from S to ${\rm P}^{\prime}$, and $\widehat{n}_{\rm s^{\prime} p}$ from ${\rm S}^{\prime}$ to ${\rm P}$. Note that both $\widehat{n}_{\rm s p^{\prime}}$ and $\widehat{n}_{\rm s^{\prime} p}$ are parallel to the Galactic plane while $\widehat{n}_{\rm s p}$ makes an angle b with it. $\lambda$ is the angle between the lines ${\rm S \, P^{\prime}}$ and ${\rm C \, P}^{\prime}$ in both of the panels. This is also the angle between ${\rm S^{\prime} \, P}$ and ${\rm C^{\prime} \, P}$ in panel-b. The direction of the vectors $\vec{a}_{\rm p, Galpl}$, $\vec{a}_{p^{\prime}, Galpl}$, and $\vec{a}_{\rm s, Galpl}$ are shown with arrows in panel-a (see text for definition of these vectors). Following the notations used in the text, ${\rm CS}=R_{\rm s}$, ${\rm C^{\prime} P} = R_{\rm p}$, and ${\rm C P^{\prime} } = R_{\rm p^{\prime}}$. It is obvious that $R_{\rm p} =R_{\rm p^{\prime}}$.}
\label{fig:schematic}
\end{figure*}

The first term in the right hand side of Eqn. (\ref{eq:doppler3}) or its equivalent in terms of the spin period and its derivative is:

\begin{equation}
 \left( \frac{\dot{P}_{\rm b} }{ {P}_{\rm b}} \right)_{\rm excess, Galpl} =   \left( \frac{\dot{P}_{\rm s} }{ {P}_{\rm s}} \right)_{\rm excess, Galpl}=  \frac{1}{c} \, ( \vec{a}_{\rm p, Galpl}  -  \vec{a}_{\rm s, Galpl} ) \cdot \widehat{n}_{\rm s p}   ~ ,
\label{eq:galpl1}
\end{equation} where $\vec{a}_{\rm p, Galpl}$ and $\vec{a}_{\rm s, Galpl}$ are the components of the accelerations of the pulsar and the sun respectively parallel to the Galactic plane, and ${a}_{\rm s, Galpl}$ and ${a}_{\rm p, Galpl}$ are the magnitudes of these vectors. These terms are usually calculated with the help of the observed rotation curve of the Galaxy. To understand this, let us assume a pulsar P, having Galactic longitude $l$ and latitude  $b$, is located at a distance $d$ from the sun, as shown in Fig. \ref{fig:schematic}. The figure also shows the Galactic center C and the sun S on the Galactic plane. If the location of the pulsar (P) is projected on the Galactic plane at ${\rm P}^{\prime}$, then $\widehat{n}_{\rm s p^{\prime}}$ becomes the unit vector from S to ${\rm P}^{\prime}$. ${\rm P^{\prime \prime}}$ is the projection of ${\rm P^{\prime}}$ on SC. Additionally, in the right panel (panel-b), ${\rm S}^{\prime}$ is the projection of S on a plane parallel to the Galactic plane at a height $z$ of the pulsar, ${\rm C}^{\prime}$ is the projection of C on the same plane. We define $\widehat{n}_{\rm s^{\prime} p}$ as the unit vector from ${\rm S}^{\prime}$ to P and $\lambda$ as the angle between the lines ${\rm S^{\prime} \, P}$ and ${\rm C^{\prime} \, P}$. Naturally, $\lambda$ is also the angle between ${\rm S \, P^{\prime}}$ and ${\rm C \, P^{\prime}}$.

As the Milky Way is in a stable configuration, $\vec{a}_{\rm s, Galpl}$ and $ \vec{a}_{\rm p, Galpl}$ can be identified with the equilibrium centripetal accelerations at S and P respectively. Similarly, the equilibrium centripetal acceleration at ${\rm P}^{\prime}$ can be denoted by $\vec{a}_{\rm p^{\prime}, Galpl}$ whose magnitude is ${a}_{\rm p^{\prime}, Galpl}$. So, we can write:

\begin{subequations}
\label{eq:centrepetalaccelerationsAll}
\begin{equation}
{a}_{\rm s, Galpl} = \frac{v_{\rm s}^2}{R_{\rm s}} ~,
\label{eq:centrepetalaccelerationsS}
\end{equation}
\begin{equation}
{a}_{\rm p, Galpl} = \frac{v_{\rm p}^2}{R_{\rm p}} ~,
\label{eq:centrepetalaccelerationsP}
\end{equation}
\begin{equation}
 {a}_{\rm p^{\prime}, Galpl} = \frac{v_{\rm p^{\prime}}^2}{R_{p^{\prime}}} = \frac{v_{\rm p^{\prime}}^2}{R_{\rm p}} ~,
\label{eq:centrepetalaccelerationsPprime}
\end{equation}
\end{subequations} where $R_{\rm s}$ ($\sim 8$ kpc) is the Galactocentric cylindrical radius of the sun, $R_p$ is the Galactocentric cylindrical radius at P and $R_{p^{\prime}} $ is the Galactocentric cylindrical radius at ${\rm P}^{\prime}$. The choice of the cylindrical coordinate system makes $R_p = R_{p^{\prime}}$ and from now on we will simply use the symbol $R_p$. In the above equations,  $v_{\rm s}$ is the Galactic rotational speed at the location of the sun, $v_{\rm p}$ is the Galactic rotational speed at ${\rm P}$, and $v_{\rm p^{\prime}}$ is the Galactic rotational speed at ${\rm P}^{\prime}$. It is obvious that the magnitude of $ \vec{a}_{\rm p^{\prime}, Galpl}$ is constant at a fixed value of $R_{\rm p}$, but its direction depends on the value of $l$ as it is always directed towards C. Similarly, $ \vec{a}_{\rm p, Galpl}$ is always directed towards ${\rm C^{\prime}}$. Neglecting the height (perpendicular to the Galactic plane) dependence of the Galactic rotation curve, one usually assumes $v_{\rm p} = v_{\rm p^{\prime}}$ giving $\vec{a}_{\rm p, Galpl} = \vec{a}_{\rm p^{\prime}, Galpl}$. This approximation gives

\begin{equation}
\left( \frac{\dot{P}_{\rm b} }{ {P}_{\rm b}} \right)_{\rm excess, Galpl}   = \left( \frac{\dot{P}_{\rm s} }{ {P}_{\rm s}} \right)_{\rm excess, Galpl}  = \frac{ 1 }{c} \, \widehat{n}_{\rm s p^{\prime}} \cdot (\vec{a}_{\rm p^{\prime}, Galpl} - \vec{a}_{\rm s, Galpl}) \, \cos b ~.
\label{eq:excessGalR1a}
\end{equation}

Fig. \ref{subfig:schematic1} makes it obvious that $ \vec{a}_{\rm s, Galpl} \cdot  \widehat{n}_{\rm s p^{\prime}} = {a}_{\rm s, Galpl} \, \cos l$ and $  \vec{a}_{\rm p^{\prime}, Galpl} \cdot  \widehat{n}_{\rm s p^{\prime}}  = {a}_{\rm p^{\prime}, Galpl} \, \cos(\pi-\lambda)= - {a}_{\rm p^{\prime}, Galpl}  \,\cos \lambda$. Using these expressions with Eqns. (\ref{eq:centrepetalaccelerationsS}) and (\ref{eq:centrepetalaccelerationsPprime}), Eqn. (\ref{eq:excessGalR1a}) simplifies to:

\begin{equation}
\left( \frac{\dot{P}_{\rm b} }{ {P}_{\rm b}} \right)_{\rm excess, Galpl}  = \left( \frac{\dot{P}_{\rm s} }{ {P}_{\rm s}} \right)_{\rm excess, Galpl} = -\frac{1}{c}  \left( {a}_{\rm p^{\prime}, Galpl}  \cos \lambda + {a}_{\rm s, Galpl} \cos l \right) \, \cos b = -\frac{1}{c}  \left(  \frac{v_{\rm p^{\prime}}^2}{R_{\rm p}} \cos \lambda + \frac{v_{\rm s}^2}{R_{\rm s}} \cos l \right) \, \cos b ~.
\label{eq:excessGalR1}
\end{equation}

To eliminate the unknown angle $\lambda$ from Eqn. (\ref{eq:excessGalR1}), one can use the triangle law as (see Fig. \ref{subfig:schematic1}):

\begin{subequations}
\begin{equation}
R_{\rm p}^2 = R_{\rm s}^2 + (d \cos b)^2 - 2 R_{\rm s} (d \cos b) \cos l ~ ,
\label{eq:Rpprime}
\end{equation} 
\begin{equation}
R_{\rm s}^2 = R_{\rm p}^2 + (d \cos b)^2 - 2 R_{\rm p} (d \cos b) \cos \lambda ~.
\label{eq:Rs1}
\end{equation} 
\end{subequations} From Eqns. (\ref{eq:Rpprime}) and (\ref{eq:Rs1}), one gets $\cos \lambda = \frac{R_{\rm s}}{R_{\rm p}} \left( \frac{d \cos b}{R_{\rm s}} - \cos l\right) = \beta \frac{R_{\rm s}}{R_{\rm p}} $  where $\beta= \frac{d \cos b}{R_{\rm s}} - \cos l$ and $ \frac{ R_{\rm p}^2 }{ R_{\rm s}^2 } = \sin^2 l + \beta^2$. So, Eqn. (\ref{eq:excessGalR1}) reduces to: 

\begin{equation}
\left( \frac{\dot{P}_{\rm b} }{ {P}_{\rm b}} \right)_{\rm excess, Galpl} = \left( \frac{\dot{P}_{\rm s} }{ {P}_{\rm s}} \right)_{\rm excess, Galpl}  = -\frac{1}{c} \frac{v_{\rm s}^2}{R_{\rm s}} \left(  \cos l + \frac{v_{\rm p^{\prime}}^2}{v_{\rm s}^2} \, \frac{\beta}{(\sin^2 l + \beta^2)} \right) \, \cos b ~.
\label{eq:excessGalR2}
\end{equation} It is obvious that the sign of $\left( \frac{\dot{P}_{\rm b} }{ {P}_{\rm b}} \right)_{\rm excess, Galpl}$ depends on the values of $l$, $b$, $d$, as well as on the form of the Galactic rotation curve. Eqn. (\ref{eq:excessGalR2}) is the exact expression of $\left( \frac{\dot{P}_{\rm b} }{ {P}_{\rm b}} \right)_{\rm excess, Galpl}$ provided the assumption $v_{\rm p^{\prime}} = v_{\rm p}$ holds valid.

As the majority of known pulsars are located near the solar system ($R_{\rm p} \sim R_{\rm s}$), it is a common practice to use a linear form for the Galactic rotation curve, as:

 \begin{equation}
v_{\rm p^{\prime}}  =  v_{\rm s} +  \left. \frac{dv}{dR} \right|_{R=R_{\rm s}} \left( R_{\rm p} - R_{\rm s} \right) = v_{\rm s} \left( 1 - b_0 \, \frac{R_{\rm p} - R_{\rm s}}{R_{\rm s}}  \right) ~,
\label{eq:linearrot}
\end{equation} where $b_0 = -  \left. \frac{R_{\rm s}}{v_{\rm s}} \frac{dv}{dR}  \right|_{R=R_{\rm s}} $ is known as the `slope parameter'. \citet{dt91} used Eqn. (\ref{eq:linearrot}) with a negligibly small value of $b_0 =  0.00 \pm 0.03$, i.e., $v_{\rm p^{\prime}} \simeq v_{\rm s}$ and Eqn. (\ref{eq:excessGalR2}) with $\cos b = 1 $ ($b \simeq 0^{\circ}$) to obtain

\begin{equation}
\left( \frac{\dot{P}_{\rm b} }{ {P}_{\rm b}} \right)_{\rm excess, Galpl, b=0} = \left( \frac{\dot{P}_{\rm s} }{ {P}_{\rm s}} \right)_{\rm excess, Galpl, b=0} = -\frac{1}{c} \frac{v_{\rm s}^2}{R_{\rm s}} \left(  \cos l +  \frac{\beta}{(\sin^2 l + \beta^2)} \right) ~.
\label{eq:excessGaldt91}
\end{equation} 

If we remove the assumption $\cos b = 1 $, but keep $v_{\rm p^{\prime}}  \simeq v_{\rm s}$, we can write

\begin{equation}
\left( \frac{\dot{P}_{\rm b} }{ {P}_{\rm b}} \right)_{\rm excess, Galpl} = \left( \frac{\dot{P}_{\rm s} }{ {P}_{\rm s}} \right)_{\rm excess, Galpl}  = -\frac{1}{c} \frac{v_{\rm s}^2}{R_{\rm s}} \left(  \cos l +  \frac{\beta}{(\sin^2 l + \beta^2)} \right)\, \cos b~.
\label{eq:excessGaldt91present}
\end{equation}

Often, people simply use Eqn. (\ref{eq:excessGaldt91present}) with recently estimated values of the parameters, e.g., $R_{\rm s} = 8.34 \pm 0.16 \,{\rm kpc}$ and $v_{\rm s} = 240 \pm 8 \,{\rm km ~ s^{-1}}$ \citep{rmb14, mth16}. However, \citet{rmb14} also obtained $  \left. \frac{dv}{dR} \right|_{R=R_{\rm s}} = - 0.2 \pm 0.4 ~{\rm km ~ s^{-1} ~ kpc^{-1}}$. This rotation curve is also almost flat, but not exactly as the one chosen by \citet{dt91}, so it would be more logical to use Eqn. (\ref{eq:linearrot}) with these updated values of the parameters to calculate $v_{\rm p^{\prime}}$ and then use that value in Eqn. (\ref{eq:excessGalR2}) to estimate $\left( \frac{\dot{P}_{\rm b} }{ {P}_{\rm b}} \right)_{\rm excess, Galpl}$.

Nonetheless, Eqn. (\ref{eq:linearrot}) is not valid at all for pulsars with $R_{\rm p} < 4 $ kpc where the Galactic rotation curve is not linear at all \citep{cgl18}. In such a case one can replace Eqn. (\ref{eq:linearrot}) by a more realistic rotation curve, say the one returned by `galpy'. `galpy' is a publicly available package\footnote{https://github.com/jobovy/galpy} modeling the gravitational potential of the Galaxy, which we have decided to use and will discuss in details later. The value of $v_{\rm p^{\prime}}$ from that rotation curve can be used in Eqn. (\ref{eq:excessGalR2})  to estimate the value of  $\left( \frac{\dot{P}_{\rm b} }{ {P}_{\rm b}} \right)_{\rm excess, Galpl}$. However, for high-latitude pulsars, the assumption of $v_{\rm p} = v_{\rm p^{\prime}}$ resulting in Eqn. (\ref{eq:excessGalR2}) should not be used, and one should rather find a better way to estimate $\left( \frac{\dot{P}_{\rm b} }{ {P}_{\rm b}} \right)_{\rm excess, Galpl}$.

To obtain a better expression for $\left( \frac{\dot{P}_{\rm b} }{ {P}_{\rm b}} \right)_{\rm excess, Galpl}$, let us look at Fig. \ref{subfig:schematic2}, which  shows that $ \vec{a}_{\rm s, Galpl} \cdot  \widehat{n}_{\rm s p} = ( \vec{a}_{\rm s, Galpl} \cdot  \widehat{n}_{\rm s p^{\prime}}) \cos b   = {a}_{\rm s, Galpl} \, \cos l \cos b $ and $ \vec{a}_{\rm p, Galpl} \cdot  \widehat{n}_{\rm s p} = ( \vec{a}_{\rm p, Galpl} \cdot  \widehat{n}_{\rm s^{\prime} p}) \cos b  = {a}_{\rm p, Galpl} \, \cos (\pi - \lambda) \cos b = - {a}_{\rm p, Galpl}  \cos \lambda \cos b$. So, Eqn. (\ref{eq:galpl1}) becomes

\begin{equation}
\left( \frac{\dot{P}_{\rm b} }{ {P}_{\rm b}} \right)_{\rm excess, Galpl}  = -\frac{1}{c}  \left( {a}_{\rm p, Galpl} \, \cos \lambda + {a}_{\rm s, Galpl} \, \cos l \right)\, \cos b ~. 
\label{eq:excessGal2018}
\end{equation} Note the difference between Eqn. (\ref{eq:excessGalR1}) and Eqn. (\ref{eq:excessGal2018}) - the first term in  Eqn. (\ref{eq:excessGalR1}) is $ {a}_{\rm p^{\prime}, Galpl} \, \cos \lambda$, while it is $ {a}_{\rm p, Galpl} \, \cos \lambda$ in  Eqn. (\ref{eq:excessGal2018}). This difference will be important for high latitude pulsars, for which one can not simply assume ${a}_{\rm p, Galpl} = {a}_{\rm p^{\prime}, Galpl}$. So, one needs to calculate the values of ${a}_{\rm p, Galpl} $ and $ {a}_{\rm s, Galpl} $ accurately to get a precise estimate of $\left( \frac{\dot{P}_{\rm b} }{ {P}_{\rm b}} \right)_{\rm excess, Galpl}$. We opt to perform these tasks using a good model of the Galactic potential, provided by galpy. The usual expression  $\cos \lambda = \frac{R_{\rm s}}{R_{\rm p}} \left( \frac{d \cos b}{R_{\rm s}} - \cos l\right) $ is used here, too. 

\subsubsection{Contribution from the relative acceleration perpendicular to the Galactic plane}
\label{subsec:galz}

\begin{figure*}
\begin{center}
\includegraphics[width=0.50\textwidth]{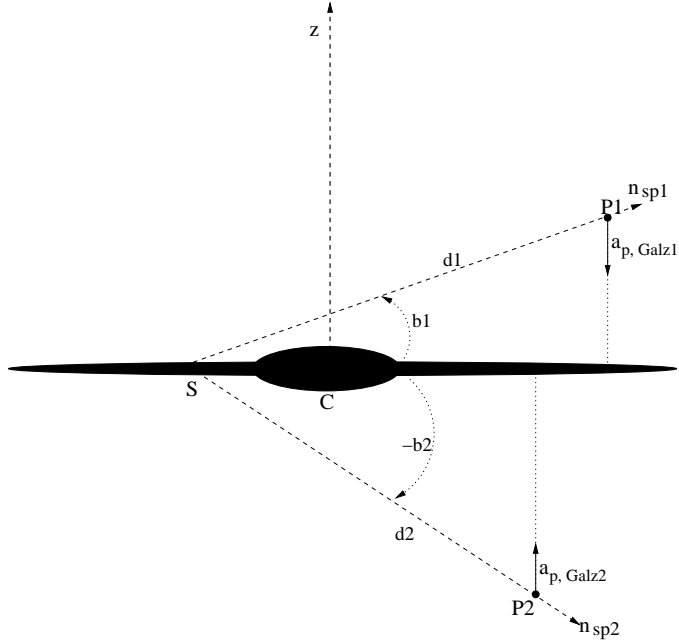}
\end{center}
\caption{Schematic diagram to understand the $z$-acceleration. Two pulsars, P1 at positive $z_1,~b_1$ and P2 at negative $z_2,~b_2$ are shown. S is the location of the sun and C is the Galactic center.}
\label{fig:schematic2}
\end{figure*}

Similarly, we can write

\begin{equation}
\left( \frac{\dot{P}_{\rm b} }{ {P}_{\rm b}} \right)_{\rm excess, Galz}  =  \left( \frac{\dot{P}_{\rm s} }{ {P}_{\rm s}} \right)_{\rm excess, Galz} = \frac{ 1 }{c} \, \widehat{n}_{\rm s p} \cdot (\vec{a}_{\rm p, Galz} - \vec{a}_{\rm s, Galz}) = \frac{ 1 }{c} \, \widehat{n}_{\rm s p} \cdot \vec{a}_{\rm p, Galz} = - \frac{ 1 }{c} \, | {a}_{\rm p, Galz} | \, \sin |b| ~,
\label{eq:excessGalz1}
\end{equation} where $\vec{a}_{\rm p, Galz}$ and $\vec{a}_{\rm s, Galz}$ are the components of the accelerations of the pulsar and the sun respectively perpendicular to the Galactic plane and the magnitudes of these two vectors are ${a}_{\rm p, Galz}$ and ${a}_{\rm s, Galz}$ respectively. We have assumed $\vec{a}_{\rm s, Galz}= 0 $, as the sun is located in the Galactic plane. The vertical height is expressed as $z = d \, \sin b$, i.e., positive for positive values of $b$ and negative for negative values of $b$. This means that the sign of $z$ bears the sense of the direction. Note that, as $\vec{a}_{\rm p, Galz}$ is directed downward (negative) for positive values of $z$, $b$ and directed upward (positive) for negative values of $z$, $b$, i.e., $\widehat{n}_{\rm s p} \cdot \vec{a}_{\rm p, Galz}$ is always negative (See Fig. \ref{fig:schematic2}).

Interestingly, there are many efforts to evaluate this perpendicular component of the acceleration, commonly known as $K_z$, due to the gravitational field of the Galaxy. One usually first determines the densities of different components of the Galaxy using observed positions, velocities, numbers and luminosities of different types of stars (mainly K-giants), and then solves Poisson's equation for each component to find the value of $K_z$ \citep[and references therein]{hf00}. Such works are usually done for `local' stars, i.e., for stars very close to the sun. \citet{nt95} used one such work \citep{kg89} when they were studying PSR J2019$+$2425 ($R_p \simeq 7.92$ kpc, $z \simeq -0.13$ kpc) and PSR J2322$+$2057 ($R_p \simeq 8.47$ kpc, $z \simeq -0.61$ kpc), as both of these pulsars are close enough to the sun. The analytical expression for ${a}_{\rm p, Galz}/c$ (or $K_z/c$) used by \citet{nt95} can be written (with the direction of the vectors for better clarity) as:

\begin{equation}
\frac{\vec{a}_{\rm p, Galz}}{c}  = - 1.08\times10^{-19} \left[0.58 + \frac{1.25}{(z_{\rm kpc}^2 + 0.0324)^{1/2}}  \right] \, z_{\rm kpc}  ~~ {\rm s^{-1}} ~ \widehat{z} ~,
\label{eq:nt4}
\end{equation} where $z_{\rm kpc}$ is the vertical height of the pulsar in the unit of kpc and $\widehat{z}$ is a vertically upward unit vector (along the positive z direction).  Afterwards, this expression has been used for other pulsars, even for the ones that are not that close to the sun. Recently, \citet{dcl16} used revised values of $K_z$ by \citet{hf04}. The $z$ dependence of $K_z$ is shown in the Figure 8 of \citet{hf04}, which we fit with a functional form as:

\begin{eqnarray}
\frac{\vec{a}_{\rm p, Galz}}{c}  =  -  1.08101 \times10^{-19}\left[0.47 + \frac{1.56 }{({{z}^{2}_{\rm kpc}} + 0.1673)^{1/2}} + \frac{0.01 }{({{z}^{2}_{\rm kpc}} + 0.0318)^{3/2}}\right] \, {z_{\rm kpc}} ~~ {\rm s^{-1}} ~~~ \widehat{z} ~~~~ {\rm for} ~ |z_{kpc}| \leq 1.5  \nonumber \\
= - 1.08101 \times10^{-19}\left[0.54 + \frac{1.43}{({{z}^{2}_{\rm kpc}} + 0.0467)^{1/2}}\right] \, {z_{\rm kpc}} ~~ {\rm s^{-1}} ~~~ \widehat{z} ~~~~ {\rm for} ~ |z_{kpc}| > 1.5  ~. \nonumber \\ 
\label{eq:dhruvfitHF04}
\end{eqnarray} In both of the Eqns. (\ref{eq:nt4}) and (\ref{eq:dhruvfitHF04}), the multiplicative factor $1.08\times10^{-19}$ arises to convert ${\rm km^2 \, s^{-2} \, pc^{-1}/} c$ to the unit of ${\rm s^{-1}}$, and $z_{\rm kpc} = d_{\rm kpc} \, \sin b$ is negative for negative values of $b$, where $d_{\rm kpc}$, the distance of the pulsar from the sun in the unit of kpc, is always positive. The minus sign before the numerical factor is used to make $\vec{a}_{\rm p, Galz} / c$ positive for negative values of $z_{\rm kpc}$ (and vice versa) as expected.

\begin{figure*}
\begin{center}
\subfigure[]{\label{subfig:hf04fit}\includegraphics[width=0.49\textwidth,angle=-0]{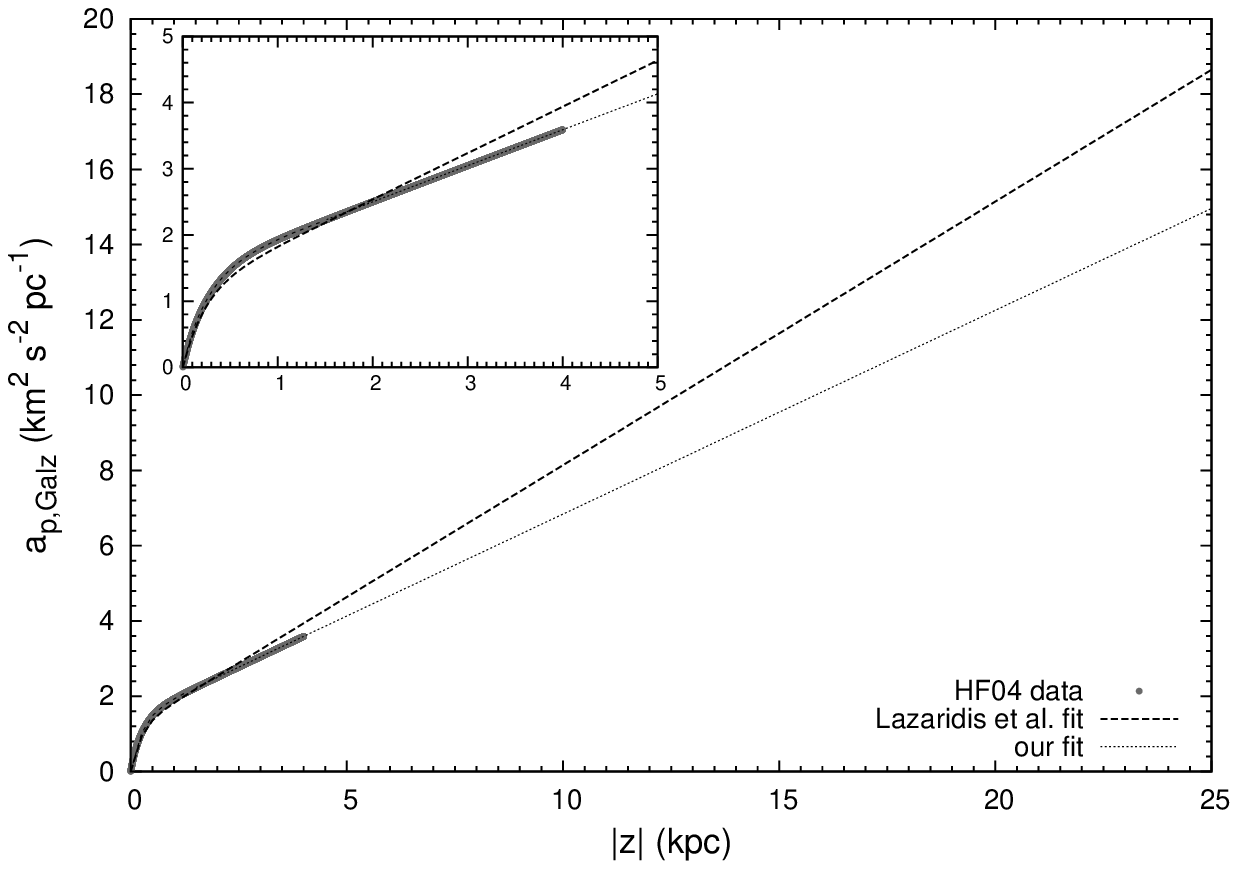}}
\subfigure[]{\label{subfig:hf04NT}\includegraphics[width=0.49\textwidth,angle=0]{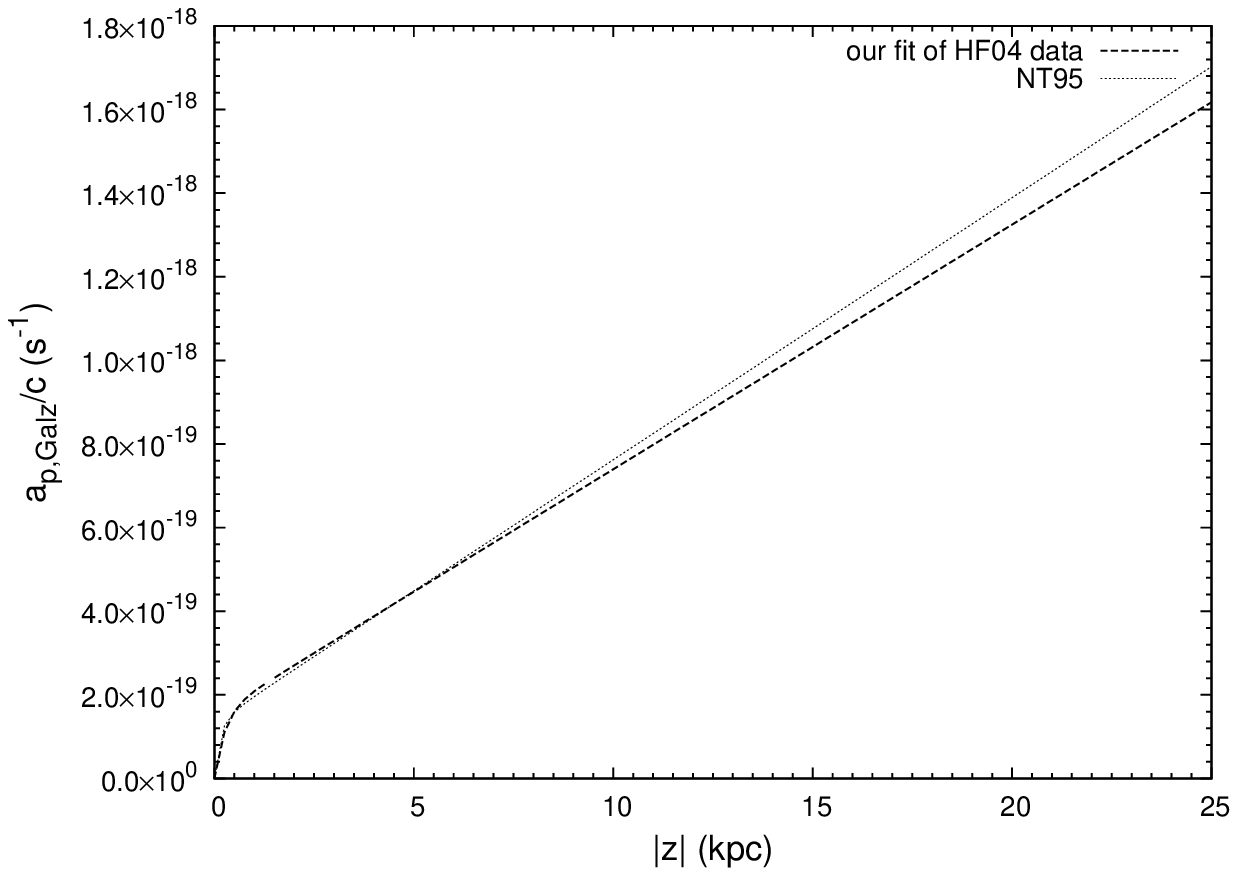}}
\end{center}
\caption{The vertical acceleration due the gravitational potential of the Galaxy. In both of the panels, the absolute value of the vertical height in kpc (${\rm |z_{kpc}|}$) is shown along the abscissa. The left panel (panel-a) compares our fit (the dotted line) as in Eqn. (\ref{eq:dhruvfitHF04}), and the fit by \citet{lwz09} (the dashed line) with the data of \citet{hf04} (dark circles plotted upto ${ |z_{\rm kpc}|} \sim 5$). We also show a zoomed in (upto ${ |z_{\rm kpc}|} = 5$) plot in the inset. Note the mixed unit along the ordinate that represents ${a}_{\rm p, Galz}$. A multiplicative factor of $3.24078 \times 10^{-11}$ will convert this into the SI unit ${\rm m \, s^{-2}}$. The right panel (panel-b) compares our fit (Eqn. (\ref{eq:dhruvfitHF04}), the dashed line) with the expression given by \citet{nt95} (Eqn. \ref{eq:nt4}, the dotted line). A multiplicative factor of $3.24078 \times 10^{-11}/c = 1.08101 \times 10^{-19} $ has been used to obtain the values of ${a}_{\rm p, Galz}/c$ in ${\rm s^{-1}}$.}
\label{fig:hf04fitNT}
\end{figure*}

In the left panel of Fig \ref{fig:hf04fitNT}, we show our fit (Eqn. \ref{eq:dhruvfitHF04}) and the earlier fit by \citet{lwz09} (the dashed line) with the data of \citet{hf04} (dark circles plotted upto ${ |z_{\rm kpc}|} \sim 5$). We also show a zoomed in (upto ${ |z_{\rm kpc}|} = 5$) plot in the inset. The discrepancy between the two fits for $|z| > 5 ~{\rm kpc}$ is clearly visible. Moreover, even at low values of $|z|$, i.e. $|z| < 1.5 ~{\rm kpc}$, our fit matches the data better than the fit by \citet{lwz09}. In the right panel of Fig \ref{fig:hf04fitNT}, we compare Eqns. (\ref{eq:nt4}) and (\ref{eq:dhruvfitHF04}). It is clear that for upto $z \sim 10$ kpc, both of these equations give almost the same results. However, even \citet{hf04} used only the stars close to the sun. So, it will not be accurate enough to even use Eqn. (\ref{eq:dhruvfitHF04}) for pulsars with $R_{\rm p}$ significantly different than $R_{\rm s}$ and/or very high values of $|z|$. In such cases, it would be preferable to adopt a better method to calculate the value of ${a}_{\rm p, Galz} / c$. More recently, \citet[figure 17]{br13} gave a $K_z(R)$ law for stars in somewhat larger range of the Galactocentric radius ($R$), i.e., between  5 to 9 kpc, but keeping $z=1.1$ kpc. So, even this work is not suitable to use for pulsars in any arbitrary locations in the Galaxy, and we again decide to use the gravitational potential of the Galaxy provided by galpy to resolve this issue.

Intuitively, at a very high value of $|z|$, a pulsar would experience less gravitational force, hence ${a}_{\rm p, Galz}  / c $ should start decreasing with the increase of $|z|$ after a certain value of $|z|$. Both Eqns. (\ref{eq:nt4}) and (\ref{eq:dhruvfitHF04}) fail to hint for this trend. The realistic potential used for the Galaxy in galpy reveals this feature.

\subsection{Dynamical effect due to the motion of the pulsar relative to the sun or the Shklovskii effect}
\label{subsec:shk}

The second term in the right hand side of Eqn. (\ref{eq:doppler2b}), which involves the relative velocity and the change of the location of the pulsar, is the well-known `Shklovskii term' \citep{shk70} and is denoted by $\left( \frac{\dot{P}_{\rm b} }{ {P}_{\rm b}} \right)_{\rm excess, Shk}$ or $\left( \frac{\dot{P}_{\rm s} }{ {P}_{\rm s}} \right)_{\rm excess, Shk}$. Thus,

\begin{equation}
\left( \frac{\dot{P}_{\rm b} }{ {P}_{\rm b}} \right)_{\rm excess, Shk} = \left( \frac{\dot{P}_{\rm s} }{ {P}_{\rm s}} \right)_{\rm excess, Shk} = -   \frac{1}{c} (\vec{V}_{\rm s} - \vec{V}_{\rm p}) \cdot \frac{d}{dt} (\widehat{n}_{\rm s p}) =    \frac{1}{c} \vec{V}_{\rm ps} \cdot \frac{d}{dt} (\widehat{n}_{\rm s p}) = \frac{1}{c} ( \dot{d} \, \widehat{n}_{sp}  + d \, \dot{\theta} \widehat{n}_{T}) \cdot (\dot{\theta} \, \widehat{n}_{T} ) \, ,
\label{eq:excessSh1}
\end{equation} where $ \vec{V}_{ps} = \vec{V}_p - \vec{V}_s $ is the velocity of the pulsar relative to the sun. One can decompose it into two orthogonal components, one is along to the line-of-sight, i.e., the radial velocity $\vec{V}_{d} = \dot{d} \, \widehat{n}_{sp} $ and the other one is perpendicular to the line-of-sight, i.e., the transverse velocity $\vec{V}_{T} = d \, \dot{\theta} \, \widehat{n}_{T}$ where $\theta$ is the two dimensional polar angle, i.e., the angle $\widehat{n}_{sp}$ makes with a reference direction in a plane containing $\vec{V}_{d}$ and $\vec{V}_{T}$. Here $ \widehat{n}_{s p}$ and $ \widehat{n}_{T}$ are unit vectors along the line-of-sight and perpendicular to it  respectively, so that $d \widehat{n}_{sp}/dt = \dot{\theta} \, \widehat{n}_{T} $, $d \widehat{n}_{T}/dt = - \dot{\theta} \, \widehat{n}_{sp}$, and $ \widehat{n}_{sp} \cdot \widehat{n}_{T} = 0$. So, we get

\begin{equation}
\left( \frac{\dot{P}_{\rm b} }{ {P}_{\rm b}} \right)_{\rm excess, Shk} = \left( \frac{\dot{P}_{\rm s} }{ {P}_{\rm s}} \right)_{\rm excess, Shk} =  \frac{1}{c} d \dot{\theta}^2 = \frac{1}{c} \frac{V_T^2}{d} =  2.42925 \times 10^{-21}  \, d_{\rm kpc} \, \mu_{T, {\rm mas \, yr^{-1}}}^2 ~ ~{\rm s^{-1}} \, .
\label{eq:excessSh2}
\end{equation} 

The transverse motion of a celestial object is usually measured in terms of angular coordinates, and is called the `proper motion'. The transverse speed ($V_T$) is related to the total proper motion ($\mu_T$) as $ V_T = d \, \mu_T$. In Eqn. (\ref{eq:excessSh2}), $\mu_{T, {\rm mas \, yr^{-1}}}$ is the total proper motion of the pulsar relative to the sun, measured in the unit of milliarcsecond per year. The total proper motion can be written as $\mu_T = \sqrt{\mu_{\delta}^2+\mu_{\alpha}^2}$, where $\mu_{\alpha}$ is the proper motion in the right ascension ($\alpha$), and $\mu_{\delta}$ is the proper motion in the declination ($\delta$). Sometimes, the proper motion of a pulsar is fitted in terms of its ecliptic coordinates, and in such a case $\mu_T = \sqrt{\mu_{e\lambda}^2+\mu_{e\beta}^2}$, where $\mu_{e\lambda}$ is the proper motion in the ecliptic longitude ($e\lambda$), and $\mu_{e\beta}$ is the proper motion in the ecliptic latitude ($e\beta$).

Note that, $\left( \frac{\dot{P}_{\rm b} }{ {P}_{\rm b}} \right)_{\rm excess, Shk} = \left( \frac{\dot{P}_{\rm s} }{ {P}_{\rm s}} \right)_{\rm excess, Shk}$ is always positive. For most of the pulsars, this is the largest dynamical contribution and easy to calculate if accurate enough values of the distance and the proper motion of the pulsar are known.

\subsection{Additional dynamical contributions when the pulsar is located in a globular cluster}
\label{subsec:gc}

In the regions where the local gravitational potential is large enough, there will be extra dynamical effects. Globular clusters are the best examples of such regions. For a pulsar located in a globular cluster, there will be two additional terms, the first is the acceleration ($\vec{a}_{\rm p, gc}$) of the pulsar due to the overall potential of the cluster and the second one is the acceleration ($\vec{a}_{\rm p, stars}$) of the pulsar due to the gravitational potential of one or more very close by stars in the same cluster \citep{bra87, phin92, phin93}. $\vec{a}_{\rm p, stars}$ is usually very small \citep{phin92, prf17} and its effect is likely to be perceptible only in the higher order derivatives of the period \citep{frk17}. $\vec{a}_{\rm p, gc} $ depends not only on the location of the pulsar inside the globular cluster, but also on the mass distribution of the cluster.  One needs to evaluate this term with a well measured location of the pulsar and a good model for the host globular cluster as recently done by \citet{frk17} for pulsars in 47 Tucanae and by \citet{prf17} for pulsars in Terzan 5. 

Moreover, one needs to use the high precision proper motions of the pulsars measured through timing analysis to estimate $\left( \frac{\dot{P}_{\rm b} }{ {P}_{\rm b}} \right)_{\rm excess, Shk} = \left( \frac{\dot{P}_{\rm s} }{ {P}_{\rm s}} \right)_{\rm excess, Shk}$, as the overall proper motion of the cluster as seen by optical astrometric instruments like \textit{Hipparcos} or \textit{Gaia} will be different than the proper motions of the pulsars \citep{frk17} in that cluster.

Note that the higher order (time derivative) dynamical effects involving jerks, jounces etc. are also likely to be significant inside a globular cluster as there the  gradient of the gravitational potential changes rapidly with the spatial coordinates and a slight movement of the pulsar from one position to another yields a change in its acceleration. Moreover, in such dense stellar environments, the probability of close fly-bys causing a change in the acceleration of the pulsar is also high.

Finally, if a pulsar is close to the center of the cluster and orbits around the center (or a massive object at the center), and this orbital motion is unmodeled, then the pulsar will experience another extra effect due to the line-of-sight component of the orbital acceleration as will be discussed in section \ref{subsec:orbitalacceleration}. By modeling this effect, \citet{bsl17} concluded the existence of an intermediate-mass black hole at the center of the globular cluster NGC 6624. 

\subsection{The contribution of the orbital acceleration in $\left( \frac{\dot{P}_{\rm s} }{ {P}_{\rm s}} \right)_{\rm excess}$}
\label{subsec:orbitalacceleration}

We have seen that both the rate of change of the orbital and the spin periods are affected by the dynamical factors in a similar fashion. Additionally, if the pulsar is a member of an unmodeled binary (usually in a very wide orbit, so that a good timing solution can be obtained even without fitting for binary parameters), there would be an additional term due to the line-of-sight component of the orbital acceleration of the pulsar:

\begin{equation}
\left( \frac{\dot{P}_{\rm s} }{ {P}_{\rm s}} \right)_{\rm excess, orbit}  =  \frac{ 1 }{c} \, \widehat{n}_{\rm s p} \cdot \vec{a}_{\rm p, orbit} =   \frac{ 1 }{c} {a}_{\rm {p}_{l, orbit}} ~.
\label{eq:excessOrbit}
\end{equation} The expression for the line-of-sight component of the orbital acceleration, ${a}_{\rm {p}_{l, orbit}}$ can be found in the literature, one of those being \citet{blw13}. 

Higher order derivatives that depend on the line-of-sight component of the jerk, jounce, etc., might also be significant, depending on the properties of the binary. \citet{jr97} gave the simple expression:

\begin{equation}
\left. \frac{1}{{P}_{\rm s}} \frac{d^k \, P_{\rm s}}{dt^k} \right|_{\rm excess, orbit}  = \frac{ 1 }{c} \frac{d^{k-1} \, {a}_{\rm {p}_{l, orbit} } }{dt^{k-1}} ~, ~~~~~  k \geq 2 ~.
\label{eq:excessOrbit2}
\end{equation}

The values of ${a}_{\rm {p}_{l, orbit}}$ and its derivatives depend on the properties of the binary not known a priori, e.g., the masses of the components, the size of the orbit, the eccentricity, the orientation of the orbit, etc. However, it is possible to constrain the allowed ranges of the parameters using the values of $\left. \frac{1}{{P}_{\rm s}} \frac{d^k \, P_{\rm s}}{dt^k} \right|_{\rm excess, orbit} $.

\section{The gravitational potential of the Galaxy as modeled in galpy}
\label{sec:galpy}

As we have already mentioned that we have decided to use the publicly available model of the Galaxy - galpy, to estimate the values of ${a}_{\rm {p}, Galpl} / c$ and ${a}_{\rm {p}, Galz} / c$ directly, here we discuss this model briefly. More details can be found in \citet{bovy15}. The default potential for the Milky Way in galpy is a combination of three potentials, a Miyamoto-Nagai disk potential \citep{mn75}, a spherical power-law density with an exponential cut-off to model the potential of the Galactic bulge, and a Navarro-Frenk-White potential \citep{NFW97} for the dark matter halo. The resulting potential $\Phi_{\rm MW}(R, |z|)$ is a function of the Galactocentric cylindrical radius $R$ and the absolute value of the vertical height $|z|$ and is known as {\tt MWPotential2014} in galpy. However, this default potential does not include the super-massive black hole at the Galactic center. It is suggested to include the effect of the super-massive black hole by adding a Kepler potential with a proper choice of the mass of the black hole (e.g., $4 \times 10^6 \, {\rm M_{\odot}}$) with {\tt MWPotential2014}.

In galpy, ${a}_{\rm p, Galpl}$ is known as ${\tt Rforce}$ and ${a}_{\rm p, Galz}$ as ${\tt zforce}$. As {\tt Rforce} $=   - \frac{\partial\,\Phi_{\rm MW}(R, |z|)}{\partial\,R}$ and {\tt zforce} $=  - \frac{\partial\,\Phi_{\rm MW}(R, |z|)}{\partial\,z}$, {\tt Rforce} is always negative and {\tt zforce} is negative for positive $z$ and vice versa. Moreover, both of them are functions of both $R$ and $|z|$. As galpy works best in its natural units where all distances are in the units of $R_{\rm s}$ and velocities are in the unit of $v_{\rm s}$, one needs to perform proper unit conversions to use galpy functions.

\begin{figure*}
\begin{center}
\subfigure[]{\label{subfig:Rforceplot}\includegraphics[width=0.34\textwidth,angle=-90]{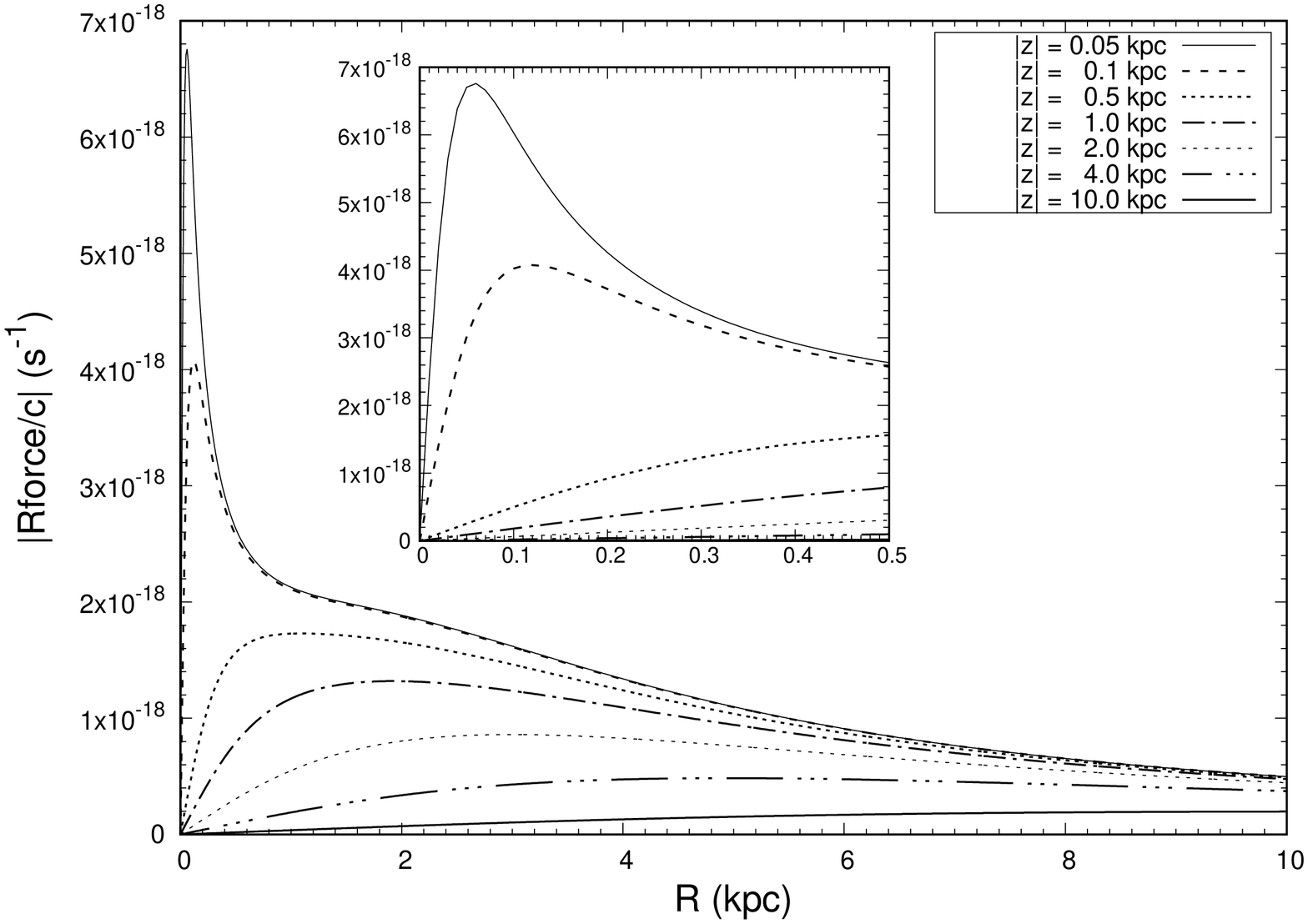}}
\subfigure[]{\label{subfig:Zforceplot}\includegraphics[width=0.34\textwidth,angle=-90]{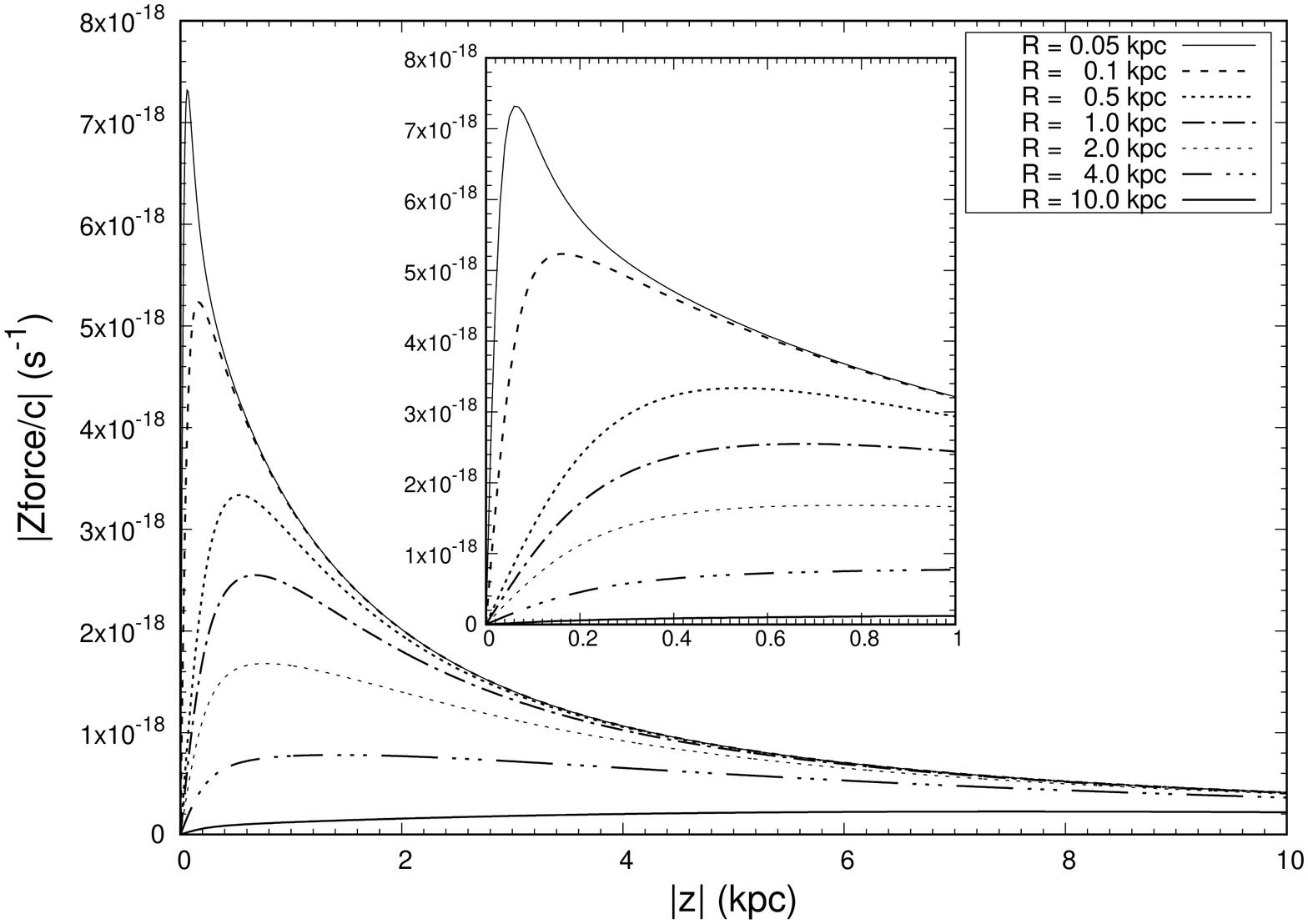}}
\end{center}
\caption{The left panel shows the variation of $|{\tt Rforce}|/c$ with $R$ for different fixed values of $|z|$. The inset shows the low $R$ region zoomed in. The right panel shows the variation of $|{\tt zforce}|/c$ with $|z|$ for different fixed values of $R$. The inset shows the low $|z|$ region zoomed in. For both of the panels, galpy's default potential for the Galaxy without the central super-massive black-hole {\tt MWPotential2014} has been used with $R_{\rm s}= 8.0~{\rm kpc}$ and $v_{\rm s} = 220~{\rm km \, s^{-1}}$.}
\label{fig:RforceZforceplot}
\end{figure*}

By plotting ${\tt |Rforce|}/c$ against $R$ for different fixed values of $|z|$ (the left panel of Fig. \ref{fig:RforceZforceplot}), we find that it first increases and then decreases with the increase of $R$. The slope of the curve is steeper at smaller $|z|$ in both of the rising and the falling sides. For any fixed $|z|$, the rise is much steeper than the fall. ${\tt |Rforce|}/c$ reaches its peak value at $R < 1 $ kpc unless $|z| > 1$ kpc. The value of ${\tt |Rforce|}/c$ is larger at smaller values of $|z|$ for any particular value of $R$. This difference is maximum near the peak value. At large $R$, curves are almost flat.  To obtain the above plot, galpy's default potential for the Galaxy without the central super-massive black-hole, i.e., {\tt MWPotential2014} has been used with $R_{\rm s}= 8.0~{\rm kpc}$ and $v_{\rm s} = 220~{\rm km \, s^{-1}}$.

Similarly, using the same potential, we plot ${\tt |zforce|}/c$ against $|z|$ for different fixed values of $R$ in the right panel of Fig. \ref{fig:RforceZforceplot}.  We find that it first increases and then decreases with the increase of $|z|$.  The slope of the curve is steeper at smaller $R$ in both of the rising and the falling sides. For each $R$, the rise is steeper than the fall. ${\tt |zforce|}/c$ reaches its peak value at $|z| < 0.5 $ kpc. The value of ${\tt |zforce|}/c$ is larger at lower values of $R$ for any particular value of $|z|$. This difference is maximum near the peak value. At large $|z|$, curves are almost flat. Comparison between Fig. \ref{subfig:hf04NT} and Fig. \ref{subfig:Zforceplot} shows a clear difference between the conventional and galpy produced values and $|z|$ dependence of ${a}_{\rm {p}, Galz} / c$.

Note that, $\left( \frac{\dot{P}_{\rm b} }{ {P}_{\rm b}} \right)_{\rm excess, Galz}$ depends not only on $|a_{\rm p, Galz}|$, i.e., ${\tt |zforce|}$, but also on $\sin |b|$ (see Eqn. (\ref{eq:excessGalz1})), so at very low values of $|b|$, one will get a low value of $\left( \frac{\dot{P}_{\rm b} }{ {P}_{\rm b}} \right)_{\rm excess, Galz}$ even if $a_{\rm p, Galz}$ is high (possible at very low values of $R$). Similarly, Eqn. (\ref{eq:excessGal2018}) shows that  $\left( \frac{\dot{P}_{\rm b} }{ {P}_{\rm b}} \right)_{\rm excess, Galpl}$ depends not only on $a_{\rm p, Galpl}$ but also on $a_{\rm s, Galpl}$, $\cos \lambda$, and $\cos l$.

We have already mentioned that galpy has the option of using the Galactic potential without or with the black hole. However, the addition of the black hole does not make much change, varying both $R$ and $z$ over the range of $0.01 - 10.0$ kpc, we find that both the ratio of the {\tt Rforce} with the black hole to that without the black hole and the ratio of the {\tt zforce} with the black hole to that without the black hole remains less than 1.36. However, as expected, these ratios can be large at very small values of $R$ and $|z|$. As an example, at $R = z = 0.001$ kpc, both of these ratios become 4.81. 

However, in the region so close to the Galactic center, there will be additional sources of acceleration of the pulsars, e.g., molecular gases (especially in the central molecular zone in $l$ ranging from  $-0^{\circ}.7 {\rm ~ to~} 1^{\circ}.7$, and $b$ ranging from  $-0^{\circ}.2 {\rm ~ to~} 0^{\circ}.2$), nearby stars, etc. These effects will depend on the exact location of the pulsar.

Another important point to remember is the fact that the default values of the parameters in galpy are $R_{\rm s}= 8.0~{\rm kpc}$ and $v_{\rm s} = 220~{\rm km \, s^{-1}}$ (defined under names `ro' and `vo' respectively in a file `\$home/.galpyrc'), which have been used to fit various observational data \citep[section 3.5]{bovy15}. These values agree with the recent conclusion by \citet{cdr18} that the best choice is $R_{\rm s}= 8.0~ \pm 0.17 ~{\rm kpc}$ and $v_{\rm s} = 220~ \pm 7 ~{\rm km \, s^{-1}}$ (both $1\sigma$ errors). One can in principle change these parameters by editing the file `\$home/.galpyrc', but in such a case one must fit other parameters of the potential too as explained in \citet{bovy15}.

\section{Improvements in the methods of estimation of dynamical effects: introduction to GalDynPsr}
\label{sec:galdynpsrIntro}

Here we discuss our package GalDynPsr that can estimate various dynamical terms more accurately than conventional methods using a model of the gravitational potential of the Galaxy provided by the galpy package. Note that GalDynPsr does not calculate special dynamical terms, e.g., the one arising due to a local potential, the one due to an unmodeled orbital motion, etc.

Various models presently available in GalDynPsr are listed in Table \ref{tb:modelsinGalDynPsr}. Some of the models (A, B, C, D) follow the conventional approaches as discussed in sections \ref{subsec:galpl} and \ref{subsec:galz}, which would be fine for near-by (within 1 kpc distance from the sun) pulsars.  Model-L estimates the values of ${a}_{\rm p, Galpl} / c$, ${a}_{\rm {s}, Galpl} / c$ and ${a}_{\rm p, Galz} /c$ using galpy. There are some semi-conventional models (G, I, K), where instead of Eqn. (\ref{eq:linearrot}), GalDynPsr calculates $v_{\rm p^{\prime}}$ using the rotation curve returned by galpy, and use this value of $v_{\rm p^{\prime}}$ in Eqn. (\ref{eq:excessGalR2}) to estimate the value of $\left( \frac{\dot{P}_{\rm b} }{ {P}_{\rm b}} \right)_{\rm excess, Galpl}$. However, at high $|z|$ values $v_{\rm p} \ne v_{\rm p^{\prime}}$ and Eqn. (\ref{eq:excessGalR2}) is not valid (see the derivation of Eqn. (\ref{eq:excessGalR2}) in section \ref{subsec:galpl}).  So this method of calculating $\left( \frac{\dot{P}_{\rm b} }{ {P}_{\rm b}} \right)_{\rm excess, Galpl}$ is also inaccurate for high $|z|$ pulsars. GalDynPsr also has a few mixed models (E, F, H, J), where either one of $\left( \frac{\dot{P}_{\rm b} }{ {P}_{\rm b}} \right)_{\rm excess, Galpl}$ and $\left( \frac{\dot{P}_{\rm b} }{ {P}_{\rm b}} \right)_{\rm excess, Galz}$ is calculated using a conventional method and the other one by using galpy. Each model involving galpy has two sub-classes, denoted by `a' and `b'. An `a' in the name of the model means that the model uses galpy without adding the contribution of the central super-massive black hole in the Milky Way potential, while a `b' in the name of the model means that the model uses galpy after adding the contribution of the central super-massive black hole in the Milky Way potential (as already discussed). GalDynPsr can also calculate $\left( \frac{\dot{P}_{\rm b} }{ {P}_{\rm b}} \right)_{\rm excess, Shk}$ if the proper motion of the pulsar is known. It also has the option of calculating the dynamical terms for pulsars in globular clusters using the cluster parameters provided in a file made using the catalogue by W. Harris \citep[2010 edition available at {\tt http://physwww.physics.mcmaster.ca/$\sim$harris/mwgc.dat}]{harris96} and assuming those to be the values of the parameters of the pulsar itself. GaldynPsr has two versions, one is a standalone script (GaldynPsrScript, available at {\tt https://github.com/pathakdhruv/GalDynPsrScript$\_$py3}) where inputs are to be given as command line arguments. The second version of GalDynPsr is designed to be usable as a library (importable module) by other python programs which would provide the values of the parameters needed. The second version is available at {\tt https://github.com/pathakdhruv/GalDynPsr}, and can be even installed using the {\tt pip3} command of python. This version is also available at http://doi.org/10.5281/zenodo.1461551. Both of these versions can return the values of the fractional and the absolute dynamical terms for any model. The first version is more useful when the user wants to estimate the dynamical terms for a single pulsar, and the second version is more useful when the user wants to use these results as a part of bigger calculations or simulations. Details of usage are available in the package documentation. One needs basic python libraries, like `scipy', `numpy', `astropy', and `galpy' to be installed in the system to use GalDynPsr.

GalDynPsr returns uncertainties in dynamical terms except when it uses globular cluster parameters from the Harris catalogue or when it uses galpy. It performs standard error propagation calculations using user provided values of the uncertainties in $l$, $b$, $d$, $\mu_{\alpha}$, and $\mu_{\delta}$. During a pulsar timing analysis, one usually fits right acsension and declination very accurately and then converts those values to $l$ and $b$. The value of $d$ is usually the least accurate, even when estimated from the parallax. GalDynPsr also reads the uncertainties in $R_{\rm s}$, $v_{\rm s}$, and $\left. \frac{dv}{dR} \right|_{R=R_{\rm s}}$ (or $b_0$) from a parameter file provided with the package. The user can change the values in this file if they wish.

\begin{table}
\caption{Models available in GalDynPsr. The columns from the left to the right are the name of the model, the method of estimating $\left( \frac{\dot{P}_{\rm b} }{ {P}_{\rm b}} \right)_{\rm excess, Galpl} = \left( \frac{\dot{P}_{\rm s} }{ {P}_{\rm s}} \right)_{\rm excess, Galpl}$, and the method of estimating $\left( \frac{\dot{P}_{\rm b} }{ {P}_{\rm b}} \right)_{\rm excess, Galz} = \left( \frac{\dot{P}_{\rm s} }{ {P}_{\rm s}} \right)_{\rm excess, Galz}$. Each model involving galpy has two options: (a) without the super-massive black hole (BH) and (b) with the super-massive black hole. For each of the models, users have the freedom to change the values of the parameters involved (see text for details).}
\begin{tabular}{l l l l}
\hline \hline
 & Method of calculating  & & Method of calculating \\ 
 & $\left( \frac{\dot{P}_{\rm b} }{ {P}_{\rm b}} \right)_{\rm excess, Galpl} = \left( \frac{\dot{P}_{\rm s} }{ {P}_{\rm s}} \right)_{\rm excess, Galpl}$ & & $\left( \frac{\dot{P}_{\rm b} }{ {P}_{\rm b}} \right)_{\rm excess, Galz} = \left( \frac{\dot{P}_{\rm s} }{ {P}_{\rm s}} \right)_{\rm excess, Galz}$ \\ \hline
Model-A & Eqn. (\ref{eq:excessGaldt91present}) & &   Eqn. (\ref{eq:nt4})   \\ \\
Model-B & Eqn. (\ref{eq:excessGaldt91present}) &  &  Eqn. (\ref{eq:dhruvfitHF04}) \\ \\
Model-C & Eqns. (\ref{eq:Rpprime}, \ref{eq:linearrot}, \ref{eq:excessGalR2}) & &  Eqn. (\ref{eq:nt4}) \\ \\
Model-D & Eqns. (\ref{eq:Rpprime}, \ref{eq:linearrot}, \ref{eq:excessGalR2}) & & Eqn. (\ref{eq:dhruvfitHF04}) \\ \\ \hline

Model-Ea & Eqn. (\ref{eq:excessGaldt91present}) & & `zforce' in galpy (without BH)  \\ \\
Model-Eb & Eqn. (\ref{eq:excessGaldt91present}) & & `zforce' in galpy (with BH) \\ \\ \hline

Model-Fa & Eqns. (\ref{eq:Rpprime}, \ref{eq:linearrot}, \ref{eq:excessGalR2})  & &  `zforce' in galpy (without BH) \\ \\
Model-Fb & Eqns. (\ref{eq:Rpprime}, \ref{eq:linearrot}, \ref{eq:excessGalR2})  & &  `zforce' in galpy (with BH) \\ \\ \hline

Model-Ga & $v_{\rm p^{\prime}} / v_{\rm s}$ `galpy' (without BH) + Eqn. (\ref{eq:excessGalR2}) & &  Eqn. (\ref{eq:nt4})  \\ \\
Model-Gb & $v_{\rm p^{\prime}} / v_{\rm s}$ `galpy' (with BH) + Eqn. (\ref{eq:excessGalR2}) & &  Eqn. (\ref{eq:nt4})  \\ \\ \hline 

Model-Ha & `Rforce' in `galpy' (without BH) & &  Eqn. (\ref{eq:nt4})  \\ \\
Model-Hb & `Rforce' in `galpy' (with BH) & &  Eqn. (\ref{eq:nt4})  \\ \\ \hline

Model-Ia & $v_{\rm p^{\prime}} / v_{\rm s}$ galpy (without BH) + Eqn. (\ref{eq:excessGalR2}) & &  Eqn. (\ref{eq:dhruvfitHF04}) \\ \\
Model-Ib & $v_{\rm p^{\prime}} / v_{\rm s}$ galpy (without BH) + Eqn. (\ref{eq:excessGalR2}) & &  Eqn. (\ref{eq:dhruvfitHF04}) \\ \\ \hline 

Model-Ja &  `Rforce' in galpy (without BH) & &  Eqn. (\ref{eq:dhruvfitHF04}) \\ \\
Model-Jb &  `Rforce' in galpy (with BH) & &  Eqn. (\ref{eq:dhruvfitHF04}) \\ \\ \hline

Model-Ka & $v_{\rm p^{\prime}} / v_{\rm s}$ galpy (without BH) + Eqn. (\ref{eq:excessGalR2}) & &  `zforce' in galpy (without BH)  \\  \\
Model-Kb & $v_{\rm p^{\prime}} / v_{\rm s}$ galpy (with BH) + Eqn. (\ref{eq:excessGalR2}) & &  `zforce' in galpy (with BH) \\  \\ \hline

Model-La & `Rforce' in galpy (without BH) & &  `zforce' in galpy (without BH)  \\  \\
Model-Lb & `Rforce' in galpy (with BH) & &  `zforce' in galpy (with BH) \\  \hline \hline

\label{tb:modelsinGalDynPsr}
\end{tabular}
\end{table}

\section{Demonstration of GalDynPsr}
\label{sec:galdynpsrdemo}

First, to check the efficiency of GalDynPsr, we confirm that it agrees with the results available in the literature when we use models involving conventional methods. As an example, \citet{prf17} reported $  (\vec{a}_{\rm p} - \vec{a}_{\rm s}) \cdot \widehat{n}_{\rm s p}  = 5.1 \times 10^{-10}~ {\rm m \, s^{-2}}$ using $l=3^{\circ}.8$, $b=1^{\circ}.7$, and $d = 5.9$ kpc for Terzan 5 ($R_{\rm s}=8.34$ kpc, $v_{\rm s}= 240~{\rm km \, s^{-1}}$). This value matches with the value returned by model-A of GalDynPsr if we use the same values of $R_{\rm s}$, $v_{\rm s}$, $d$, $l$, and $b$. Model-A in GalDynPsr represents the traditional approach taken by \citet{prf17}. However, they used more recently measured value of $d$, that is smaller than the value quoted in the  Harris catalogue, i.e., $d=6.9$ kpc. We also obtain $\dot{P}_{\rm b, Gal} = - 0.008 \times 10^{-12}~ {\rm s s^{-1}}$ for PSR B1913+16 using the latest parameters reported by \citet{dwnc18} in both model-B and model-La. This value matches with the value reported by them. As this pulsar is located at a low $|z|$ values, i.e. at $z=0.15$ kpc and $R_{\rm p} = 6.21$ kpc, the conventional method is sufficient. We also find that the difference between the values of $\dot{P}_{\rm b, Gal}$ obtained in models B and La is $7 \times 10^{-17} ~ {\rm s s^{-1}}$ (calculated using the best values of the parameters and ignoring the errors). This difference is ignorable in regard to the present day accuracy of $\dot{P}_{\rm b, obs} = -2.423 \pm 0.001 \times 10^{-12} ~{\rm s \, s^{-1}}$, but will be of importance when the precision of $\dot{P}_{\rm b, obs}$ measurement will improve by two or more orders and the use of model-La will make more sense.

In Table \ref{tb:compareModelsGalDynPsrb}, we report values of various fractional dynamical terms estimated for four sample pulsars (the only ones for which \citet{dcl16} could measure $\dot{P}_{\rm b, obs}$, see their table 16) and two most pulsar populated globular clusters, Terzan 5 and 47 Tucanae, using different models of GalDynPsr. The reason behind the choice of pulsars was to be able to compare with existing estimates of dynamical terms. The parameters for the globular clusters are taken from the Harris catalogue \citep[2010 edition, for globular clusters]{harris96}. For the pulsars, $l$ and $b$ values are calculated using the {\tt SkyCoord} module of astropy from the reported values of the right ascension and the declination in tables 3, 4, and 8 of \citet{dcl16}. The values of $\mu_{\alpha}$ and $\mu_{\delta}$ are taken from the same tables. The values of $d$ are taken from table 16 of \citet{dcl16}. For the sake of simplicity, we do not report uncertainties in the table, although used while running GalDynPsr. Moreover, we display the values of the dynamical terms upto the fourth decimal places, just to compare between models. In reality, these values are precise only upto the second decimal places as the uncertainties in the values of $d$ appear in the second decimal places in \citet{dcl16}. We have used $R_{\rm s}=8.00 \pm 0.17$ kpc and $v_{\rm s}= 220 \pm 7~{\rm km \, s^{-1}}$, that are used in galpy (without the uncertainties) to fit the parameters of the Milky Way potential to agree with observations. Our main findings are summarized as:

\begin{enumerate}

\item Our model-B is the closest to the method used by \citet{dcl16}. However, we still see some discrepancies. The main reason is the fact that we used $R_{\rm s}= 8.0~ \pm 0.17 ~{\rm kpc}$ and $v_{\rm s} = 220~ \pm 7 ~{\rm km \, s^{-1}}$  \citep{cdr18} while \citet{dcl16} used  $R_{\rm s}= 8.34  ~{\rm kpc}$ and $v_{\rm s} = 240 ~{\rm km \, s^{-1}}$. Using the later set of values of $R_{\rm s}$ and $v_{\rm s}$, we obtained similar values for $ \left( \frac{\dot{P}_{\rm b} }{ {P}_{\rm b}} \right)_{\rm excess, Galpl}$, i.e., $3.30 \times 10^{-20}~ {\rm s^{-1}}$, $5.4 8\times 10^{-20}~ {\rm s^{-1}}$, $2.99 \times 10^{-20}~ {\rm s^{-1}}$, and $10.50 \times 10^{-20}~ {\rm s^{-1}}$ for pulsars J0613-0200, J0751+1807, J1012$+$5307, and J1909$-$3744 respectively. However, the use of this set of values of $R_{\rm s}$ and $v_{\rm s}$ produces little (in the third decimal places) difference in the values of  $\left( \frac{\dot{P}_{\rm b} }{ {P}_{\rm b}} \right)_{\rm excess, Galz}$ than those reported in Table \ref{tb:compareModelsGalDynPsrb}. We do not recommend the use of model-B as it uses the traditional approach with drawbacks as already discussed.

\item Our preferred model-La gives significantly different values of both $ \left( \frac{\dot{P}_{\rm b} }{ {P}_{\rm b}} \right)_{\rm excess, Galpl}$ and $ \left( \frac{\dot{P}_{\rm b} }{ {P}_{\rm b}} \right)_{\rm excess, Galz}$ from the values quoted in \citet{dcl16} for PSR J1012$-$5307, as this pulsar has a moderately high value of $|z|$.

For PSR J1909$-$3744, the values of $ \left( \frac{\dot{P}_{\rm b} }{ {P}_{\rm b}} \right)_{\rm excess, Galz}$ as obtained by our model-La and reported by \citet{dcl16} are significantly different, because for this pulsar, $R_p$ is significantly smaller than $R_s$, as well as $|z|$ is not too small.

For PSR J0613$-$0200 and PSR J0751$+$1807, model-La gives the values for $ \left( \frac{\dot{P}_{\rm b} }{ {P}_{\rm b}} \right)_{\rm excess, Galz}$ almost the same as the ones reported by \citet{dcl16}. The reason is the fact that for both of these pulsars $R_p \sim R_s$ and $|z|$ is not too high. Moreover, the low $|z|$ value of PSR J0613$-$0200 makes the value of $ \left( \frac{\dot{P}_{\rm b} }{ {P}_{\rm b}} \right)_{\rm excess, Galpl}$ as obtained by model-La to be very close to the one reported by \citet{dcl16}. On the other hand, a larger value of $|z|$ for PSR J0751$+$1807 produces somewhat larger disagreement between the value of $ \left( \frac{\dot{P}_{\rm b} }{ {P}_{\rm b}} \right)_{\rm excess, Galpl}$ as obtained by model-La with the one reported by \citet{dcl16}.

\item We have already discussed in section \ref{sec:galpy} that the addition of the black hole does not affect much in the values of the {\tt Rforce} and {\tt zforce}, and as a result, does not affect much in the values of the dynamical terms. This fact is clear from the closeness of the results obtained with model-La and model-Lb in Table \ref{tb:compareModelsGalDynPsrb}. We still keep both of the options available in GalDynPsr.

\item When $R_{\rm p} << R_{\rm s}$, {\tt zforce} gives significantly different result than other methods. This is the case for Terzan 5, where $R_{\rm p} = 1.21$ kpc. Models involving Eqn. (\ref{eq:nt4}) give $ \left( \frac{\dot{P}_{\rm b} }{ {P}_{\rm b}} \right)_{\rm excess, Galz} = -0.34 \times  10^{-20} \, {\rm   s^{-1} }$, models involving Eqn. (\ref{eq:dhruvfitHF04}) give $ \left( \frac{\dot{P}_{\rm b} }{ {P}_{\rm b}} \right)_{\rm excess, Galz} = -0.28 \times  10^{-20} \, {\rm   s^{-1} }$, but models involving {\tt zforce} give $ \left( \frac{\dot{P}_{\rm b} }{ {P}_{\rm b}} \right)_{\rm excess, Galz} = - 4.65 \times  10^{-20} \, {\rm   s^{-1} }$. The discrepancies between different methods of calculating $ \left( \frac{\dot{P}_{\rm b} }{ {P}_{\rm b}} \right)_{\rm excess, Galz}$ are not so severe for any other examples chosen, as none of those have $R_{\rm p}$ so small. 

In such a case ($R_{{p}} << R_{\rm s}$), traditional methods also fail to give a correct value of $\left( \frac{\dot{P}_{\rm b} }{ {P}_{\rm b}} \right)_{\rm excess, Galpl}$. Again, we can take Terzan 5 as an example, and see that $\left( \frac{\dot{P}_{\rm b} }{ {P}_{\rm b}} \right)_{\rm excess, Galpl} = 322.29 \times 10^{-20} \, {\rm   s^{-1} }$ for models involving Eq. (\ref{eq:excessGaldt91present}), and as expected, the substitution of the perfect flat rotation curve of \citet{dt91} by the rotation cure of \citet{rmb14}, i.e., the use of the models involving Eqns. (\ref{eq:Rpprime}), (\ref{eq:linearrot}), and (\ref{eq:excessGalR2}) does not make much difference, gives $\left( \frac{\dot{P}_{\rm b} }{ {P}_{\rm b}} \right)_{\rm excess, Galpl} = 327.08 \times 10^{-20} \, {\rm   s^{-1} } $. On the other hand, instead of Eqn. (\ref{eq:linearrot}), if we use the rotation curve from galpy in Eqn. (\ref{eq:excessGalR2}), we get a significantly different value, i.e., $\left( \frac{\dot{P}_{\rm b} }{ {P}_{\rm b}} \right)_{\rm excess, Galpl} = 119.67 \times 10^{-20} \, {\rm   s^{-1} } $ (without the black hole), even though one still uses the approximation $v_{\rm p^{\prime}} = v_{\rm p}$. The use of {\tt Rforce} option of galpy accounts for the height dependence and hence improves the accuracy further. It gives $\left( \frac{\dot{P}_{\rm b} }{ {P}_{\rm b}} \right)_{\rm excess, Galpl} = 112.98 \times 10^{-20} \, {\rm   s^{-1} } $ without the black hole. 

The above mentioned points establish the fact that for a pulsar with $R_{\rm p} << R_{\rm s}$, one must abandon conventional methods to calculate the dynamical terms, and opt for a more modern method like model-La or model-Lb provided in GalDynPsr. 

\item The role of a high value of $|z|$ in these dynamical terms will be clear from a comparison between PSR J0613$-$0200 and PSR J1012$+$5307. These two pulsars have almost the same value of $R_{\rm p}$, i.e. $8.67$ kpc and $8.69$ kpc respectively, but somewhat different value of $|z|$, $0.13$ kpc and $0.89$ kpc respectively. We see larger difference between the value of $\left( \frac{\dot{P}_{\rm b} }{ {P}_{\rm b}} \right)_{\rm excess, Galz}$ obtained using {\tt zforce} and that obtained using conventional methods for PSR J1012$+$5307 which has larger $|z|$. Similarly, this pulsar shows a larger disagreement in the values of $\left( \frac{\dot{P}_{\rm b} }{ {P}_{\rm b}} \right)_{\rm excess, Galpl}$ obtained using {\tt Rforce} and that obtained using conventional methods.

\item In addition to the comparison between models as demonstrated in Table 2, we have also explored the difference between the fractional dynamical contributions from the Galactic potential, i.e., $\left( \frac{\dot{P}_{\rm b} }{ {P}_{\rm b}} \right)_{\rm excess, Gal}$ as returned by the full conventional method (model-A) and the full galpy based model (model-La) over the full range of $l$ and $b$ for different values of $d$. The difference increases with the increase of $d$, and when $d \ge 1$ kpc, the absolute value of the difference becomes greater than $ 2 \times 10^{-20}~ {\rm s^{-1}}$ for most of the $l-b$ phase space. As the value of $\left( \frac{\dot{P}_{\rm b} }{ {P}_{\rm b}} \right)_{\rm excess, Gal}$ is usually in the order of $10^{-20}~ {\rm s^{-1}}$ (see Table \ref{tb:compareModelsGalDynPsrb} for some examples), we can conclude that the accuracy of the model can impact the result if $d \ge 1$ kpc.

\item GalDynPsr does not return uncertainties in the fractional dynamical terms when it uses models involving galpy, because galpy does not return uncertainties. However, because of the modular structure of GalDynPsr, one can estimate uncertainties by emplyoing Monte-Carlo simulation. As an example, for  PSR J0613$-$0200, model-La returns $\left( \frac{\dot{P}_{\rm b} }{ {P}_{\rm b}} \right)_{\rm excess, Galpl} = 4.00 ~ \times {\rm  10^{-20} \, s^{-1} }$ and $\left( \frac{\dot{P}_{\rm b} }{ {P}_{\rm b}} \right)_{\rm excess, Galz} = -9.29 ~ \times {\rm  10^{-21} \, s^{-1} }$ using the values of $l$, $b$, and $d$ as given in Table \ref{tb:compareModelsGalDynPsrb}. We then simulated 50000 instances of $l$, $b$, and $d$. For $l$ and $b$, as usual we converted the mean values and uncertainties in the right ascension and the declination from table 3 of \citet{dcl16} using the {\tt SkyCoord} module of the astropy package. The mean value and the uncertainty in $d$ are taken from table 16 of \citet{dcl16}. We then calculated the fractional dynamical terms using model-La for each instance, and then calculated the mean and the standard deviation of 50000 values of $\left( \frac{\dot{P}_{\rm b} }{ {P}_{\rm b}} \right)_{\rm excess, Galpl}$ and $\left( \frac{\dot{P}_{\rm b} }{ {P}_{\rm b}} \right)_{\rm excess, Galz}$. In this manner, we found $\left( \frac{\dot{P}_{\rm b} }{ {P}_{\rm b}} \right)_{\rm excess, Galpl} = (3.99 \pm 0.41 )  \times {\rm  10^{-20} \, s^{-1} }$ and $\left( \frac{\dot{P}_{\rm b} }{ {P}_{\rm b}} \right)_{\rm excess, Galz} = - (9.25 \pm 0.62) \times {\rm  10^{-21} \, s^{-1} }$ respectively.

\end{enumerate}

\begin{table}
\caption{Comparison of the values of the fractional dynamical terms estimated using different methods for the pulsars mentioned in table 16 of \citet{dcl16}, as well as for the two most pulsar populated globular clusters, Terzan 5 and 47 Tucanae. The parameters for the globular clusters are taken from the Harris catalogue \citep[2010 edition, for globular clusters]{harris96}. For the 4 pulsars, $l$, $b$ values are calculated using the {\tt SkyCoord} module of astropy from the reported values of the right ascension and the declination in tables 3, 4, and 8 of \citet{dcl16}. The values of $\mu_{\alpha}$ and $\mu_{\delta}$ can be found in the same tables. The values of $d$ are taken from table 16 of \citet{dcl16}. For the sake of simplicity, we do not report uncertainties here, although used while running GalDynPsr. Moreover, we display many significant digits for the sake of comparison between models. We have used $R_{\rm s}=8.00 \pm 0.17$ kpc and $v_{\rm s}= 220 \pm 7~{\rm km \, s^{-1}}$, that are used in galpy to fit the parameters of the Milky Way potential to agree with the observational data. }
\vskip -0.1cm
\begin{tabular}{l c c c c c c c} \hline \hline
\multicolumn{2}{l} {Pulsars/Cluster}  & PSR J0613$-$0200 & PSR J0751$+$1807 & PSR J1012$+$5307 & PSR J1909$-$3744 & Ter5 & 47Tuc\\
\hline
 \multicolumn{2}{l}{ l (deg)} & $210.4131 $ & $202.7297 $ & $160.3471 $ & $359.7308$ &  3.84  & 305.89     \\
 \multicolumn{2}{l}{ b (deg)} & $-9.3049$ & $21.0858$ & $50.8578$ & $-19.5958$ & 1.69 & $-44.89$ \\
 \multicolumn{2}{l}{ d (kpc)} & $0.78$ & $1.07$ & $1.15$ & $1.15$ & 6.9 &  4.5 \\ 
 \multicolumn{2}{l}{ $\mu_{\alpha}$ (${\rm mas ~ yr^{-1}}$)} & $1.822$  & $-2.73$ & $2.609$ & $-9.519$ & $-$ & $-$  \\ 
 \multicolumn{2}{l}{ $\mu_{\delta}$ (${\rm mas ~ yr^{-1}}$) }& $-10.355$ & $-13.4$ & $-25.482$ & $-35.775$ & $-$ & $-$    \\

 \multicolumn{2}{l}{ $ \left( \frac{\dot{P}_{\rm b} }{ {P}_{\rm b}} \right)_{\rm obs} $ (in ${\rm  10^{-20} \, s^{-1} }$) }& 46.3538 & -153.9432 & 116.7604 & 379.6512 & $-$ & $-$  \\ 
 \multicolumn{2}{l}{ $ \left( \frac{\dot{P}_{\rm s} }{ {P}_{\rm s}} \right)_{\rm obs} $ (in ${\rm  10^{-20} \, s^{-1} }$) }& 313.2142 & 223.8549 & 325.8774 & 475.8964 & $-$ & $-$  \\  \\
 
 \multicolumn{2}{l}{ $z$ (kpc)} & $-0.13$ & $0.38$ & $0.89$ & $-0.39$ & $0.20$ & $-$3.18 \\  
 \multicolumn{2}{l}{ $R_{\rm p}$ (kpc)} & 8.67 & 8.93  &  8.69 &  6.92 & 1.21 & 6.65 \\      \hline 
   \multicolumn{2}{l}{ $ \left( \frac{\dot{P}_{\rm b} }{ {P}_{\rm b}} \right)_{\rm excess, Shk}  $ (in ${\rm  10^{-20} \, s^{-1} }$)} & $20.9463$ & 48.6102 & 183.3013 & 382.8569 &  $-$  &  $-$ \\  
   \multicolumn{2}{l}{ $ =  \left( \frac{\dot{P}_{\rm s} }{ {P}_{\rm s}} \right)_{\rm excess, Shk}$ } &  &  &  &  &    &   \\  \hline \hline

    &          \multicolumn{7}{l}{Values of $ \left( \frac{\dot{P}_{\rm b} }{ {P}_{\rm b}} \right)_{\rm excess, Galpl} =  \left( \frac{\dot{P}_{\rm s} }{ {P}_{\rm s}} \right)_{\rm excess, Galpl} $ (in ${\rm  10^{-20} \, s^{-1} }$) for different models: }       \\

A, B, Ea, Eb  & Eqn. (\ref{eq:excessGaldt91present})  & 3.0142 & 4.9912 & 2.7298 & 9.6504 & 322.2856 & -14.5856  \\ \\

C, D, Fa, Fb & Eqns. (\ref{eq:Rpprime}, \ref{eq:linearrot}, \ref{eq:excessGalR2}) & 3.0786 & 5.0778 & 2.7750 & 9.7908 & 327.0843  &  -14.5547 \\ \\
 
Ga, Ia, Ka & $v_{\rm p^{\prime}} / v_{\rm s}$ galpy   & 3.9031 &  6.1821  & 3.3532  & 11.5438 & 119.6713 & -14.1771   \\
  &  \multicolumn{2}{l}{ (without BH) + Eqn. (\ref{eq:excessGalR2}) }   &  &  &  &   &      \\ \\

Gb, Ib, Kb & $v_{\rm p^{\prime}} / v_{\rm s}$ galpy   & 3.9010 & 6.1800 & 3.3517 & 11.5482 & 119.7851 & -14.1764   \\
  &    \multicolumn{2}{l}{ (with BH) + Eqn. (\ref{eq:excessGalR2}) } &  &  &  &   &      \\ \\

Ha, Ja, La & `Rforce' in galpy  & 4.0004 & 6.8354 & 5.0696 & 9.9521 & 112.9790 & -18.6831  \\
  & (without BH) &  &  &  &  &   &      \\ \\

Hb, Jb, Lb & `Rforce' in galpy  & 4.0007 & 6.8359 & 5.0699 & 9.9530 & 113.0853 & -18.6838  \\ 
 & (with BH) &  &  &  &  &   &      \\  
 
  \multicolumn{2}{l}{table 16 of \citet{dcl16}} & 3.28 & 5.50 & 3.01 & 10.49 & $-$ & $-$  \\ \hline

     &       \multicolumn{7}{l}{Values of $ \left( \frac{\dot{P}_{\rm b} }{ {P}_{\rm b}} \right)_{\rm excess, Galz} =  \left( \frac{\dot{P}_{\rm s} }{ {P}_{\rm s}} \right)_{\rm excess, Galz}$ (in ${\rm  10^{-20} \, s^{-1} }$) for different models: }       \\ 

A, C, Ga  & Eqn. (\ref{eq:nt4})  & -1.3815 & -5.2721 & -14.6102 & -4.9177 & -0.3361 & -23.5742   \\ 
Gb, Ha, Hb &   &  &  &  &  &  &  \\ \\
 
B, D, Ia & Eqn. (\ref{eq:dhruvfitHF04}) & -1.1081 & -5.0601 &-15.5418  & -4.7219 & -0.2835  &  -23.9731 \\ 
Ib, Ja, Jb  &   &  &  &  &  &  &  \\ \\
 
Ea, Fa, Ka, La  &  `zforce' in galpy  & -0.9291 & -4.0218 & -12.6005 & -6.9854 & -4.6468 & -26.7781 \\
  &  (without BH) &  &  &  &  &   & \\ \\
  
Eb, Fb, Kb, Lb  &  `zforce' in galpy  & -0.9291 & -4.0218 & -12.6007 & -6.9855 & -4.6474 & -26.7792  \\
  &  (with BH) &  &  &  &  &   & \\  
  
    \multicolumn{2}{l}{table 16 of \citet{dcl16}}  & $-1.28$ &  $-4.57$ &   $-14.55$ &  $-8.24$ & $-$ & $-$  \\ \hline \hline  
  
\label{tb:compareModelsGalDynPsrb}
\end{tabular}
\end{table}

\section{Applicability of GalDynPsr}

GalDynPsr has the potential to be used in various studies involving precise timing of pulsars, we mention some of those below.

In section \ref{sec:galdynpsrdemo}, we have seen that depending on the location of a pulsar, conventional methods can give wrong values of the dynamical terms leading to wrong values of $\dot{P}_{\rm b, Gal}$. If this $\dot{P}_{\rm b, Gal}$ is comparable to $\dot{P}_{\rm b, Shk}$, this will lead to a wrong value of $\dot{P}_{\rm b, int}$. Similar logic holds for $\dot{P}_{\rm s, int}$. It is expected that the improved timing precision with SKA1-Mid \citep{ssa14} will reveal higher order post-Newtonian terms, at least for relativistic pulsars like the double pulsar or neutron star$-$black hole binaries (if discovered). VLBI using SKA1 \citep{pgr15} will also give better parallaxes, i.e., better distance estimates as well as better proper motion measurements leading to accurate estimation of $\dot{P}_{\rm b, Shk}$ and $\dot{P}_{\rm s, Shk}$. This will be useful only if accompanied by accurate estimates of $\dot{P}_{\rm b, Gal}$ and $\dot{P}_{\rm s, Gal}$, and GalDynPsr can play a crucial role in this. Below we discuss some potential areas where precise values of $\dot{P}_{\rm b, int}$ or  $\dot{P}_{\rm s, int}$ will be a necessity.

One such area will be the tests of theories of gravity. Although the most accepted gravity theory, the general relativity (GR) has passed all the tests so far, it is expected to get better limits on various alternative theories of gravity in the future, especially in the strong field regime with more precise timing solutions of pulsars. The most obvious test will be the detection of a deviation of the value of $\dot{P}_{\rm b, int}$ from that expected from GR. We have already mentioned that for a clean binary system, the emission of the gravitational waves is responsible for $\dot{P}_{\rm b, int}$ and GR allows only quadrupolar gravitational wave. The rate of change of the orbital period due to the emission of the quadrupolar gravitational waves ($\dot{P}^{Q}_{\rm b, GW}$) depends on the values of the orbital period, the orbital eccentricity, the mass of the pulsar and the mass of the companion. Thus, under GR, $\dot{P}_{\rm b, int} = \dot{P}^{Q}_{\rm b, GW}$. However, there are various theories of gravity that allow other multipoles \citep[and references therein]{bato14}. If exists, the dipolar gravitational waves become the most significant source of radiation and hence the largest contributing factor in the value of $\dot{P}_{\rm b, int}$. One can place limits on the parameter of the gravity theory by equating $\dot{P}_{\rm b, int} = \dot{P}^{D}_{\rm b, GW} + \dot{P}^{Q}_{\rm b, GW}$ (neglecting other poles of the gravitational wave emission) where $\dot{P}^{D}_{\rm b, GW}$ is the rate of change of the orbital period due to the emission of the dipolar gravitational waves. Additionally, if the value of the gravitational constant $G$ changes with time, that will also lead to a change in the orbital period, and the rate of change of the orbital period becomes $\dot{P}_{\rm b, int} = \dot{P}^{D}_{\rm b, GW} + \dot{P}^{Q}_{\rm b, GW} +  \dot{P}^{\dot{G}}_{\rm b}$, where $\dot{P}^{\dot{G}}_{\rm b}$ is the rate of change of the orbital period due to the change of the value of $G$. The expressions for different terms depend on various parameters, including the orbital period, the masses, the orbital eccentricity, the `sensitivity' of the objects etc. These expressions can be found in various literature, one example being \citet{bato14}. If one can calculate the value of $\dot{P}^{Q}_{\rm b, GW}$ from the knowledge of relevant parameters, then it will be possible to obtain a value of $ \dot{P}^{D}_{\rm b, GW}  + \dot{P}^{\dot{G}}_{\rm b}$. Recently, \citet{zdw18} placed a limit on both $\dot{G}/{G}$ and $\kappa_D$ (a characteristic parameter of the gravity theory allowing the dipolar gravitational wave emission). As such limits depend crucially on the value of $\dot{P}_{\rm b, int}$, one needs to estimate and eliminate the dynamical terms as accurately as possible. GalDynPsr will help in performing this task better, especially if such tests are done in the future with pulsars at high $z$ or at low $R_{\rm p}$ where conventional methods fail. 

Another test of GR is the test of the Strong Equivalence Principle (SEP), which is one of the main features of GR, but usually violated in alternative theories. One important aspect of SEP is the universality of the free fall (UFF). The violation of UFF in case of a binary system can be parametrized by a dimensional parameter $|\Delta|$ that quantifies the differential acceleration between the members (of different constituents) of the freely falling binary. \citet{dasc91} showed that the UFF violation can manifest into a `forced' eccentricity that depends on the value of $|\Delta|$, the masses of the members of the binary, the orbital period of the binary, and the projection of the Galactic acceleration vector at the location of the pulsar onto the orbital plane, i.e., ${a}_{\rm p, Gal, proj} $ where $ \vec{a}_{\rm p, Gal} = \vec{a}_{\rm p, Galpl} + \vec{a}_{\rm p, Galz} $. This ${a}_{\rm p, Gal, proj} $ is related to $\vec{a}_{\rm p, Gal}$ with a multiplicative function that depends on the orientation of the orbit. The measurement of the signature of this `forced' eccentricity in the timing solution of a wide-orbit low-eccentricity binary pulsar with a low-mass white-dwarf companion helps to place an upper limit on $|\Delta|$ \citep{sfl05, gsf11, zdw18}. In such efforts, GalDynPsr can be used to estimate the value of ${a}_{\rm p, Gal} $, which might be more precise than conventional methods, depending on the location of the pulsar in the Galaxy. In fact, GalDynPsr has been already used by \citet{agh18} to place the best ever limit of the non-violation of the strong equivalence principle using the pulsar PSR J0337$+$1715 in a triple system.

Even within the framework of GR, i.e., for the binaries for which $\dot{P}^{D}_{\rm b, GW}$ and $\dot{P}^{\dot{G}}_{\rm b}$ can be ignored and SEP test is not possible, e.g., for double neutron star binaries, a wrong estimate of $\dot{P}_{\rm b, int}$ will have detrimental consequences like erroneous estimates of the masses of the pulsar and the companion, especially when one excludes the most accurately measurable post-Keplerian parameter, i.e., the rate of the periastron advance, which is affected by the Lense-Thirring effect that depends on the unknown values of the moment of inertia, the orientation of spin axis, etc. \citep[and references therein]{bagchi18}. To avoid such consequences, one should use GalDynPsr, especially if the pulsar is located far from the sun where the conventional methods are inaccurate.

Similarly, there are many cases where it is very important to estimate the values of $\dot{P}_{\rm s, int}$ as precise as possible. One such case is to understand the emission of the continuous gravitational waves from pulsars due to their rotationally induced quadrupole moments. The value of $\dot{P}_{\rm s, int}$ is needed to calculate the `spin-down' limit of the strain ($h_{0}^{sd}$) of the gravitational waves emitted at a frequency of $f_{\rm gw} = 2 f_{\rm s}$ where $f_{\rm s} = 1/P_{\rm s}$ is the spin-frequency of the neutron star. The expression for $h_{0}^{sd}$ is derived under the assumption that the total spin-down energy is being lost in the form of the gravitational waves as $h_0^{sd} =  [2.5 ( G ~ I_{\rm s}  ~ |\dot{f}_{\rm s, int}|) / (c^3 ~ d^2 ~ f_{\rm s}) ]^{0.5} $ where G is the gravitational constant, $I_{\rm s}$ is the moment of inertia of the pulsar along its spin axis, and $\dot{f}_{\rm s, int}$ is the intrinsic rate of change of the spin frequency. In reality, especially in cases of rotation powered radio pulsars, only a fraction $\eta$ of the spin-down energy is converted to the gravitational energy, and $h_0 = \eta^{0.5} \, h_{0}^{sd}$. As $\eta < 1$ giving $h_0 < h_{0}^{sd}$, one can confirm the detection of gravitational waves or can place an upper limit on the value of $h_0$ only when the detection sensitivity reaches below $h_{0}^{sd}$. Although the last science run  (O1) of the advanced-LIGO-Virgo did not detect any continuous gravitational waves, it has placed the best so far $95 \%$ upper limit of $h_0$, i.e. $h_0^{95}$ for a number of pulsars and for eight of those, $h_0^{95} < h^{sd}$ \citep{aaa17}. It is expected that in the future, the detector sensitivity will surpass the spin-down limit for more pulsars, especially for the pulsars with large values of $h_0^{sd}$. The ongoing and future pulsar surveys and follow-up timing will provide more pulsars with $f_{\rm gw}$ in the LIGO range as well as large enough values of $h_{0}^{sd}$. GalDynPsr will help us calculate the values of $h_{0}^{sd}$ by providing precise values of $\dot{f}_{\rm s, int}$ for non-globular cluster pulsars, especially if the pulsars are located in the regions where conventional methods to estimate and eliminate the dynamical effects fail.

Another application of the precise estimation of the value of $\dot{P}_{\rm s, int}$ is in the study of the pulsar `death-line' \citep[and references therein]{gsl16}, i.e., in the effort to understand the radio emission mechanism. Pulsars close to the `death-line' have small values of the rate of loss of the spin down energy  \footnote{The pulsar `death-line' is a hypothetical line in the ${P}_{\rm s} - \dot{P}_{\rm s, int}$ plane defining a maximum potential drop above which the radio emission turns off \citep{cr93}. This line is very close and almost parallel to the $\dot{E}_{\rm rot} = 10^{-30}~{\rm erg \, s^{-1}}$ line \citep{ncb15}.}, i.e. $\dot{E}_{\rm rot} = 4 \pi^2 I_{\rm s} \dot{P}_{\rm s, int} / P_{\rm s}$. A larger population of low $\dot{E}_{\rm rot}$ pulsars will improve our understanding of the `death-line' better. On the other hand, a larger population will increase the probability of some of such pulsars having large enough values of $|\dot{P}_{\rm s, Gal}|$ located in the regions where the use of the conventional methods give inaccurate values of $\dot{P}_{\rm s, Gal}$ resulting in inaccurate values of $\dot{P}_{\rm s, int}$ and $\dot{E}_{\rm rot}$.

GaldynPsr can be useful even when the timing solution is not good enough, especially when the distance measurements are not as accurate as desired. In such cases, one can use the expression $\dot{P}_{\rm b, obs} - \dot{P}_{\rm b, Gal} -\dot{P}_{\rm b, GW}^{Q} = \dot{P}_{\rm b, Shk}$ or $\dot{P}_{\rm s, obs} - \dot{P}_{\rm s, Gal} - \dot{P}_{\rm s, GW}^{Q} = \dot{P}_{\rm s, Shk}$ to obtain the value of $\dot{P}_{\rm b, Shk}$ or $\dot{P}_{\rm s, Shk}$ if at least some of the post-Keplerian parameters are measured to give sufficiently accurate values of the masses of the pulsar and the companion and hence a theoretical value of $\dot{P}_{\rm b, GW}^{Q} $, neglecting the manifestation of non-quadrupolar gravitational wave emission (if exists). One can extract the value of the distance from the value of $\dot{P}_{\rm b, Shk}$ or $\dot{P}_{\rm s, Shk}$ if the proper motion is known. The value of the distance obtained in such a manner can be compared with other distance estimates (if available). Discrepancies between the values of distances estimated in different methods have the potential to reveal inaccuracies either in the timing solution or in the methods of distance estimations, or the presence of additional effects.

Finally, although we have confined our discussions to rotation powered pulsars only, GaldynPsr can be used to estimate $\dot{P}_{\rm b, Gal}$ or $\dot{P}_{\rm s, Gal}$ for any object in the Galaxy provided their distance, location, proper motion and orbital and spin periods are known.

\section{Conclusions and summary}

We have discussed various dynamical effects manifested in the values of the rate of change of the orbital and the spin period, and different methods to estimate those effects. Although the existing approximations are often sufficient for the present-day accuracy of timing solutions, one should be careful as this accuracy is getting improved for a number of millisecond pulsars thanks to the pulsar timing array efforts.
 
Keeping the above mentioned fact in mind, we have created a package GalDynPsr that evaluates different dynamical effects following the traditional as well as improved methods based on the model of the Galactic potential provided in the package galpy. GalDynPsr is publicly available and open for contributions.

The improved methods to estimate dynamical effects will be essential for pulsars far away from the solar system (either horizontally, or vertically, or both) with precise timing solutions. Even for other pulsars, as the timing accuracy improves, one should opt for more accurate estimation of dynamical effects. We recommend the use of model La or Lb, both of which calculate ${a}_{\rm p, Galpl} /c$ and ${a}_{\rm p, Galz}  /c$  directly using galpy.

However, for a pulsar very close to the Galactic center, additional contributions due to the near-by stars or the potential of the nuclear cluster (if the pulsar is in such a cluster) will be significant. So care should be taken to extract these effects in addition to other effects calculable by GalDynPsr. Similarly, for pulsars in globular clusters, the cluster potential is to be modeled out of GalDynPsr.

All the results reported in this paper used version 1.4.0 of galpy. As both galpy and GalDynPsr are being improved continuously, potential users are requested to check for new versions on the source repositories of these packages.

We would also like to remind the readers that the model of the Galactic potential as done in galpy is not the only one, many different models exist in the literature, e.g., \citet{Mc17, PDmH17}, etc. The modular structure of GalDynPsr makes sure that the interested users can incorporate new models without much difficulties. We also hope for improvement of these existing models (including the one in galpy) using the recently (25-April, 2018) released or future \textit{Gaia} data.

Finally, we would like to remind the fact that even after removing dynamical contributions from the observed rate of change of the orbital and the spin periods, there might exist an additional external contribution due to the error in modeling the time dependent change in the dispersion measure \citep{prf17} of radio pulsars. So, care should be taken regarding this issue.

\section{Acknowledgements}

The authors thank Chris Flynn for useful suggestions and sharing the data of Figure 8 of \citet{hf04} and the anonymous reviewer for useful comments on the earlier version of the manuscript.

\end{document}